\documentclass[onecolumn]{article}
\usepackage{epsfig}

\usepackage{graphicx,color}
\usepackage{mathbbol}
\usepackage{amsmath,amssymb}
\usepackage[latin9]{inputenc} 
\usepackage[sans]{dsfont}

\newtheorem{Prop}{Proposition}
\newtheorem{defin}{Definition}
\newtheorem{thm}{Theorem}
\newtheorem{cor}{Corollary}
\newtheorem{lemma}{Lemma}

\newcommand{\beq}{\begin{equation}}
\newcommand{\eeq}{\end{equation}}
\newcommand{\beqa}{\begin{eqnarray}}
\newcommand{\eeqa}{\end{eqnarray}}
\newcommand{\beqan}{\begin{eqnarray*}}
\newcommand{\eeqan}{\end{eqnarray*}}

\newcommand{\tr}[2]{\text{Tr}_{#1}\left(#2\right)}
\newcommand{\trb}[2]{\text{Tr}_{#1}\left[#2\right]}

\newcommand{\ket}[1]{| #1 \rangle}
\newcommand{\bra}[1]{\langle #1 |}

\newcommand{\inprod}[2]{\bra{#1}#2\rangle}
\newcommand{\ketbra}[1]{\ket{#1}\bra{#1}}

\newcommand{\identity}{\mathbb{I}}
\newcommand{\hilbert}{\mathcal{H}}

\newcommand{\lind}{\mathcal{L}}
\newcommand{\neigh}{\mathcal{N}}

\newcommand{\complex}{\mathbb{C}}

\newcommand{\Hi}{\mathcal{H}}
\newcommand{\BH}{{\mathfrak B}(\Hi)}
\renewcommand{\DH}{{\mathfrak D}(\Hi)}

\newcommand{\Li}{\mathcal{L}}

\newcommand{\Ei}{\mathcal{E}}
\newcommand{\Ai}{\mathcal{A}}
\newcommand{\Ti}{\mathcal{T}}
\newcommand{\Ni}{\mathcal{N}}
\newcommand{\alg}{{\rm alg}}
\newcommand{\fix}{{\rm fix}}

\newcommand{\supp}{{\rm supp}}

\renewcommand{\ker}{{\rm ker}}
\renewcommand{\span}{{\rm span}}

\renewcommand{\dim}{{\rm dim}}

\newcommand{\Tr}{{\rm Tr}}

\newcommand{\blue}{\color{black}}

\newcommand{\lv}{\color{black}}
\newcommand{\pj}{\color{black}}

\newcommand{\qed}{\hfill $\Box$ \vskip 2ex}

\begin{document}

\setcounter{page}{1}

\vspace*{0.88truein}



\vspace{-3cm}\centerline{\bf GENERAL FIXED POINTS OF QUASI-LOCAL }
\vspace*{0.035truein}
\centerline{\bf FRUSTRATION-FREE QUANTUM SEMIGROUPS:}
\vspace*{0.035truein}
\centerline{\bf FROM INVARIANCE TO STABILIZATION}

\vspace*{0.37truein}

\centerline{\footnotesize 
PETER D. JOHNSON\footnote{E-mail: peter.d.johnson@dartmouth.edu }}
\vspace*{0.015truein}
\centerline{\footnotesize\it 
Department of Physics and Astronomy, Dartmouth College, }
\baselineskip=10pt
\centerline{\footnotesize\it 
6127 Wilder Laboratory, Hanover, NH 03755, USA }
\vspace*{10pt}
\centerline{\footnotesize
FRANCESCO TICOZZI\footnote{E-mail: ticozzi@dei.unipd.it }}
\vspace*{0.015truein}
\centerline{\footnotesize\it Dipartimento di Ingegneria dell'Informazione,
Universit\`a di Padova, } 
\baselineskip=10pt
\centerline{\footnotesize\it 
via Gradenigo 6/B, 35131 Padova, Italy, \& }
\baselineskip=10pt
\centerline{\footnotesize\it 
Department of Physics and Astronomy, Dartmouth College, }
\baselineskip=10pt
\centerline{\footnotesize\it 
6127 Wilder Laboratory, Hanover, NH 03755, USA }
\vspace*{10pt}
\centerline{\footnotesize 
LORENZA VIOLA\footnote{E-mail: lorenza.viola@dartmouth.edu }}
\vspace*{0.015truein}
\centerline{\footnotesize\it 
Department of Physics and Astronomy, Dartmouth College, }
\baselineskip=10pt
\centerline{\footnotesize\it 
6127 Wilder Laboratory, Hanover, NH 03755, USA }

\vspace*{0.225truein}

\vspace*{0.21truein}

\begin{abstract}
We investigate under which conditions a mixed state on a finite-dimensional multipartite 
quantum system may be the unique, globally stable fixed point of frustration-free semigroup 
dynamics subject to specified quasi-locality constraints.  Our central result is a linear-algebraic 
necessary and sufficient condition for a generic (full-rank) target state to be \emph{frustration-free 
quasi-locally stabilizable}, along with an explicit procedure for constructing Markovian dynamics that achieve 
stabilization. If the target state is not full-rank, we establish sufficiency under an additional condition, which is 
naturally motivated by consistency with pure-state stabilization results yet provably not necessary in general. 
{Several applications are discussed, of relevance to both dissipative quantum engineering and 
information processing, and non-equilibrium quantum statistical mechanics.  
In particular, we show that a large class of graph product states (including arbitrary thermal graph states) 
as well as Gibbs states of commuting Hamiltonians are frustration-free 
stabilizable relative to natural quasi-locality constraints.
Likewise, we provide explicit examples of non-commuting Gibbs states and non-trivially entangled 
mixed states that are stabilizable despite the lack of an underlying commuting structure, albeit scalability 
to arbitrary system size remains in this case an open question.}
\end{abstract}
{}{}
\vspace*{10pt}

\baselineskip=15pt
\section{Introduction}
\subsection{Context and motivation} 

Convergence of a dynamical system to a stable equilibrium point is a hallmark of dissipative, irreversible 
behavior. In particular, rigorously characterizing the nature and stability of equilibrium states of irreversible 
{\em quantum} evolutions is a longstanding problem central to both the mathematical theory of open 
quantum systems and the foundations of quantum statistical mechanics \cite{davies,alicki-lendi}. 
In recent years, renewed interest in these issues has been fueled by the growing theoretical and 
experimental significance of techniques for quantum reservoir engineering \cite{poyatos} and 
dissipative quantum control \cite{viola-engineering} within Quantum Information Processing (QIP). 
Representative applications that benefit from {\em engineered dissipation} include  
robust quantum state preparation, with implications for steady-state entanglement 
\cite{Kraus2008,ticozzi-markovian,Kastoryano2011,didi,reeb}, 
non-equilibrium topological phases of matter \cite{ZollerReview}, and ground-state 
cooling \cite{cooling,znidaric,James}; as well as open-system quantum simulation \cite{barreiro,barreiro2},  
steady-state dissipation-driven quantum computation \cite{wolf-dissipation,zanardi}, 
dissipative quantum gadgets and autonomous quantum error correction \cite{Kastoryano2013,devoret}, 
along with quantum-limited sensing and amplification \cite{paola,clerk}.

While applications are often developed by making reference to a specific physical setting, a common  
theme is the key role played by {\em constraints}, that may restrict the allowed dynamical models and the 
extent of the available manipulations. This motivates seeking a rigorous system-theoretic framework for 
characterizing controlled open-quantum system dynamics subject to given resource constraints. 
In this work, we focus on dissipative multipartite quantum systems described by {\em time-independent 
quasi-local (QL) semigroup dynamics}, capturing the fact that, in many physically relevant  scenarios, 
both the coherent (Hamiltonian) and irreversible (Lindblad) contributions to the semigroup generator 
must act non-trivially only on finite subsets of subsystems, determined for instance by spatial lattice geometry.
The main question we address is to determine what properties an \emph{arbitrary} target state of interest must 
satisfy in order to be the \emph{unique} stationary (``fixed'') point for a given QL constraint, thereby making the 
state globally QL-stabilizable in principle, in an asymptotic sense. 

In previous work \cite{ticozzi-ql,ticozzi-steady}, we have answered this question under the assumption that 
the target state is {\em pure}, providing in particular a necessary and sufficient linear-algebraic condition for 
the latter to be stabilizable without requiring Hamiltonian dynamics. Such pure states are called purely 
Dissipatively Quasi-Locally Stabilizable (DQLS).  While restricting to a pure target state is both a natural 
and adequate first step in the context of dissipatively preparing paradigmatic entangled states of relevance 
to QIP (such as W or GHZ states), allowing for a general 
\emph{mixed} fixed-point is crucial for a number of reasons. 
On the one hand, since mixed quantum states represent the most general possibility, this is a prerequisite for 
mathematical completeness. On the other hand, from a practical standpoint, QL stabilization of a mixed state 
which is sufficiently close to an ``unreachable'' pure target may still be valuable for QIP purposes, a notable 
example being provided by thermal graph states at sufficiently low temperature \cite{Temme2014}.
Furthermore, as physical systems in thermal equilibrium are typically far from pure, characterizing mixed-state 
QL stabilization might offer insight into thermalization dynamics as occurring in Nature and on a quantum 
computer \cite{eisert}. From this point of view, a stability analysis of thermal states of QL Hamiltonians is 
directly relevant to developing efficient simulation and sampling algorithms for the quantum 
canonical ensemble, so-called ``quantum Gibbs samplers,'' as analyzed in 
\cite{Brandao2014} for commuting Hamiltonians.
 
{In the mixed-state scenario, the problem of QL stabilization involves qualitatively different features and is 
substantially more complex.}
{This is largely due to the fact that the analysis tools used in the pure-state setting do not lend 
themselves to a formulation where the invariance property of the globally 
defined target state translates directly at the level of QL generator components.  We bypass this 
difficulty by restricting to the important class of \emph{frustration-free} (FF) semigroup dynamics 
\cite{Cubitt2015,Brandao2014}, for which global invariance of a state also implies its invariance 
under each QL component.} Physically, the FF property is known to hold within 
standard derivations of Markovian semigroup dynamics, for instance based on Davies' weak coupling limit or 
``heat-bath'' approaches generalizing classical Glauber dynamics \cite{alicki-lendi,Brandao2014}.

\subsection{Summary of results}

The paper is organized of as follows. Sec. \ref{sec:prelim} is devoted to a self-contained recollection 
and setup of the required formalism. After summarizing basic facts about open quantum system dynamics 
and their associated operator $*$-algebras, we pay special attention to characterizing the structure of 
the fixed-point set that semigroup dynamics support (Sec. \ref{sub:fixed}), and to formalizing 
the relevant QL constraints (Sec. \ref{sub:ql}). The problem of interest, namely deciding if a given state 
can be stabilized by using FF dynamics, is formulated in Sec. \ref{subsec:qlsdefns}. Such states will be 
called FFQLS. We revisit there the main line of reasoning and result for DQLS pure states, as this also 
serves to motivate the restriction to the class of FF dynamics. 
{While Sec. \ref{sec:prelim} mostly reviews known results, a 
slight generalization of existing results regarding the structural characterization of fixed points for 
quantum semigroups is contained in Theorem \ref{thm:dualkernel}; conversely, Theorem 
\ref{thm:modinvariance} identifies conditions under which an operator $*$-algebra may arise as the 
fixed-point set of completely positive dynamics while preserving a desired target state.}

Sections \ref{sec:necessary} and \ref{sec:sufficiency} are the core sections of the paper,  
presenting our new results on necessary and, respectively, sufficient conditions for FFQLS. 
In order to do that, we introduce in Sec. \ref{sub:linear} the central concept of a {\em Schmidt span}, 
and a key {result (Lemma \ref{invSS})} showing how this concept allows one to construct the smallest 
eigenspace for an operator that acts non-trivially only on a factor of a tensor-product space, and 
admits a given global eigenvector. 
In Sec. \ref{sub:invariance}, by specializing this result to QL completely-positive trace-preserving generators, 
we show that requiring invariance of a target state constrains the structure of their fixed-point set: 
{not only must the kernel include the smallest *-operator subalgebra that contains the Schmidt span 
and is closed with respect to a modified product operation, but it also needs to be invariant with respect to 
the action of the modular group associated to the target state.}
Building on that, we establish in Theorem \ref{thm:neccond} a {\em necessary}  
condition that an {\em arbitrary} target state must obey in order to be FFQLS, under a specified 
locality constraint. Two implications of this result are especially worth highlighting: first, {as we formally  
show in Appendix A, in the limiting case where the target state is pure, the FFQLS notion 
naturally recovers the simpler DQLS notion.} Second, 
whereas the existence of a unique stationary state is typically assumed on the basis of 
physical properties of the generator \cite{Brandao2014,Cubitt2015,Temme2014}, our result provides 
an independent complementary criterion.
If the target state is {\em generic} (full-rank), we show in Theorem \ref{thm:suffcond} that the necessary 
condition of Theorem \ref{thm:neccond} is, in fact, also {\em sufficient}. The proof relies on a new way to 
show that a generator is FF and is constructive, yielding an explicit procedure for the synthesis of a stabilizing 
QL FF generator. Interestingly, 
this generator may be implemented in principle by {\em purely dissipative} means (no Hamiltonian 
control), establishing a further point of contact between the FFQLS and DQLS settings. 
While we {\em conjecture} that Theorem \ref{thm:suffcond} remains valid for an arbitrary  
(non-full-rank) target, our current proof requires an additional ``support condition'' (see Theorem 
\ref{thm:suffcondnorank}) which, however, need {\em not} be obeyed in general.

In Sec. \ref{sec:appl}, we provide several illustrative applications of our general mathematical 
framework, motivated by QIP and quantum statistical mechanics. 
We begin by showing that, unlike in the pure-state setting, even trivially separable 
mixed states may fail to be FFQLS -- also implying that there are {\em classical} correlations that cannot be 
generated using constrained QL resources. We go on in Sec. \ref{sub:graph} to characterize the most general 
FFQLS class of mixed states on $d$-dimensional subsystems (qudits) that may be constructed from 
a graph, by transforming an arbitrary product state with a QL quantum circuit naturally built out of it 
-- so-called {\em graph product states}. Structurally, graph states are especially simple in 
that they may be associated to a family of QL {\em commuting} ``parent Hamiltonians,'' 
which, viewed in a suitable basis, are also {\em non-interacting} (single-particle).
Stepping up in complexity, a large and important family of states also derived from 
commuting QL Hamiltonians are {\em commuting Gibbs states}, which we examine in Sec. \ref{sub:gibbs}. 
While these states are known to be FFQLS from the analysis in \cite{Brandao2014}, they serve as a prime 
example to highlight the different focus and philosophy of our system-theoretic approach: on the one hand, 
not all Gibbs states of commuting Hamiltonians need to be FFQLS, {\em if the QL is taken as an input} of the 
problem, as natural from a control-oriented perspective; on the other hand, for every commuting Gibbs state there 
exists a physically motivated QL notion (stemming, in particular, from the stabilizing ``Davies generator'' 
\cite{Brandao2014}), relative to which our necessary conditions are indeed seen to hold -- although, we
explicitly verify this statement for a specific class in one dimension (Proposition \ref{prop:gibbs}). 

{Sec. \ref{sub:nc} is devoted to exploring FFQLS in settings where {\em no manifest commuting structure} 
may be leveraged.  Specifically, we construct and analyze three main examples.  While in the pure-state 
setting, important examples exist of target states that are QL stabilizable despite being ground states of 
FF {\em non-commuting} parent Hamiltonians -- including the paradigmatic case of the spin-$1$ 
Affleck-Kennedy-Lieb-Tasaki (AKLT) Hamiltonian \cite{AKLT,Kraus2008} -- 
we further exhibit a FFQLS family of pure generalized Dicke state on qudits, for which 
no FF commuting parent Hamiltonian exists. It is, however, for mixed target states that 
our general methods best reveal their potential, as no other approach is available 
to our knowledge.  Beyond the commuting scenario,
the problem becomes substantially more complicated, both in terms of verifying the required FFQLS 
necessary conditions, and in terms of maintaining physically meaningful QL constraints. 
We find that (full-rank) Gibbs states of a transverse-field one-dimensional Ising Hamiltonian 
can be FFQLS, 
provided that the locality constraint becomes weaker (corresponding to multi-body interactions with a 
growing weight) as the system size grows -- thereby effectively limiting scalability in practice. 
As our final example, we construct 
a FFQLS family of entangled mixed states on four qubits. Beside serving as an explicit counter-example to 
the support condition mentioned above, this additionally demonstrates how ``practical  
stabilization'' of a pure state of interest (a GHZ state in this case) may be possible via stabilization 
of a mixed state arbitrarily close to the target. Final remarks conclude in Sec. \ref{sec:end}, along  
with a discussion of implications and open questions.  }

After Appendix A, which was already mentioned, 
we present in Appendix B a result on fixed states of convex combinations of 
completely positive trace-preserving maps that extends a weaker statement used in the main text
(Corollary \ref{cor:frustrationkraus}).  Appendix C collects technical results related to ``pseudo-pure'' and pure 
Dicke states discussed in Secs. \ref{failures} and \ref{Dicke}, respectively.

\section{Preliminaries}
\label{sec:prelim}

\subsection{Notation and background}
\label{sub:background}

Consider a finite-dimensional Hilbert space $\Hi,$ $\dim(\Hi)=d,$ and let ${\mathfrak B}(\Hi)$ be the set of 
linear operators on $\Hi.$  $X^\dag$ shall denote the adjoint of $X\in{\mathfrak B}(\Hi)$, with self-adjoint operators 
$X=X^\dagger$ representing physical {\em observables}. The adjoint operation corresponds to the transpose conjugate 
when applied to a matrix representation of $X$, with the simple transpose being denoted by $X^T$ and the 
entry-wise conjugation by $X^*$. 
To avoid confusion, we shall use $\identity$ to indicate the identity operator on $\BH,$ whereas ${\cal I}$ will 
indicate the identity map (or super-operator) from $\BH$ to itself. 
We shall use $X\equiv Y$ 
to say that $X$ is defined as $Y$. 

The convex subset ${\mathfrak D}(\Hi) \subset \mathfrak B(\Hi)$ of trace-one, positive-semidefinite operators, 
called {\em density operators}, is associated to physical {\em states}.  We are concerned with state changes in the 
Sch\"odinger picture between two arbitrary points in time, say $0$ and $t>0$, which are
described by a {\em completely-positive trace-preserving} (CPTP) linear map (or quantum channel) 
on $\BH$ \cite{alicki-lendi}. A map $\Ti_t$ is CP if and only if it admits an operator-sum representation,
\begin{equation}
\label{eq:kraus}
\rho(t) = \Ti_t(\rho_0 )=\sum_k M_k\rho_0 M^\dag_k,  \quad \rho_0 \in{\mathfrak D}(\Hi), 
\end{equation}
for some $\{M_k\} \subset {\mathfrak B}(\Hi)$, and is also TP if in addition $\sum_k M_k^\dag M_k=\identity.$ 
The operators $M_k$ are referred to as (Hellwig-)Kraus operators or operation elements \cite{kraus}. The operator-sum representation of a CPTP map is not unique, and new decompositions may be obtained from unitary changes of the operators $M_k$.  Dual dynamics\footnote{While from a probabilistic and operator-algebra viewpoint it would be more natural to consider the dynamics acting on the states as (pre-)dual, we follow here the standard quantum physics notation as it allows for a more direct connection with existing work as well as a more compact notation in our context. } 
~with respect to the Hilbert-Schmidt inner product on 
${\mathfrak B}(\Hi)$ (Heisenberg picture) are associated to {\em unital} CP maps 
${\cal T}^\dag$, that is, obeying the condition ${\cal T}^\dag ({\mathbb I}) = {\mathbb I}$.

A continuous one-parameter semigroup of CPTP maps $\{{\cal T}_t\}_{t\geq 0}$, with ${\cal T}_0={\cal I}$, 
characterized by the Markov composition property 
${\cal T}_t \circ {\cal T}_s = {\cal T}_{t+s}$, for all $t,s \geq 0,$ will be referred to as a 
{\em Quantum Dynamical Semigroup} (QDS) \cite{alicki-lendi}. 
We shall denote by ${\cal L}$ the corresponding semigroup generator, ${\cal T}_t = e^{{\cal L}t}$, 
with the corresponding dual QDS $\{ {\cal T}_t^\dag \}_{t\geq 0}$ being described by the generator $\Li^\dag$.  
It is well known that $\Li$ (also referred to as the ``Liouvillian'') can be always expressed in Lindblad canonical form 
\cite{Gorini1976,lindblad,heidi}, 
that is, in units where $\hbar=1$:
\begin{equation}
\label{eq:lindblad}
\dot\rho(t)= {\cal L}\left( \rho(t)\right) \equiv -i[H,\,\rho(t)]+\sum_k \Big(
L_k\rho(t)L^\dag_k-\frac{1}{2}\{L_k^\dag L_k,\,\rho(t)\} \Big),  \;\;t \geq 0,
\end{equation}
where $H=H^\dag$ is a self-adjoint operator associated with the effective Hamiltonian 
(generally resulting from the bare system Hamiltonian plus a ``Lamb shift'' term),
and the Lindblad (or noise) operators $\{L_k\}$ specify the non-Hamiltonian component of the generator, resulting in non-unitary irreversible dynamics.  Equivalently, $\lind$ defines a valid QDS generator if and only if it may be expressed in the form (see e.g. Theorem 7.1 in \cite{wolf-notes})
\begin{equation}
\lind(\rho) \equiv {\cal E} (\rho)- (\kappa \rho + \rho \kappa^\dagger), \quad \kappa \equiv i H + 
\frac{1}{2} \,{\cal E}^\dag ({\mathbb I}), 
\label{lindbladize}
\end{equation}
where ${\cal E}$ is a CP map and the anti-Hermitian part of $\kappa$ identifies the Hamiltonian operator. 

We shall denote by $\lind(H,\{L_k\})$ the QDS generator associated to Hamiltonian $H$ and noise operators $\{L_k\}.$ Throughout this paper, both $H$ and all the $L_k$ will be assumed to be time-independent, with \eqref{eq:lindblad} thus defining a {\em linear time-invariant} dynamical system. 
It is important to recall that, as for CPTP maps, the Lindblad representation is also not unique, namely, the {\em same} generator can be associated to different Hamiltonian and noise operators (see e.g. Proposition 7.4 in \cite{wolf-notes}), 
and, further to that, the separation between the Hamiltonian and the noise operators is not univocally defined 
\cite{ticozzi-ql,wolf-notes}.  Specifically, the Liouvillian is unchanged, $\lind(H,\{L_k\})=\lind(H',\{L'_k\}),$ if the new 
operators may be obtained as (i) linear combinations of $H,\{L_k\}$ and the identity, 
$L'_k = L_k + c_k {\mathbb I},$ $H' = H -({i}/{2}) \sum_k (c_k^* L_k - c_k L_k^\dag),$ with $c_k 
\in {\mathbb C}$; or (ii) unitary linear combinations, $L_k'=\sum_l u_{kl} L_l$, $H'=H$, with $U\equiv \{u_{kl}\}$ 
a unitary matrix (and the smaller set ``padded'' with zeros if needed), corresponding to a change of operator-sum 
representation for ${\cal E}$ in Eq. \eqref{lindbladize}.

We will denote a $\dag$-closed associative subalgebra {$\Ai \subseteq \BH$ generated by a set of operators $X_1,\ldots, X_k$ as $\Ai \equiv \alg\{X_1,\ldots , X_k\}.$} If $\Ti \equiv \Ti(\{M_k\})$ and $\Li \equiv \Li(H,\{L_k\})$
are a CP map and a QDS generator, then we shall let $\alg\{\Ti\} \equiv \alg\{M_k\}$
and $\alg\{\Li\} \equiv \alg\{H, L_k \},$ respectively. These algebras are invariant with respect to the change of representation in the Kraus or, respectively, Hamiltonian and Lindblad operators since, as remarked, equivalent representations are linearly related to one another.  Let $\oplus$ denote the orthogonal direct sum. It is well known that any $\dag$-closed associative subalgebra $\Ai$ of $\BH$ admits a {block-diagonal {\em Wedderburn decomposition} }
\cite{Davidson}, 
namely, $\Hi$ may be decomposed in an orthogonal sum of tensor-product bipartite subspaces, possibly up to a summand: 
\beq
\label{hildec}
\Hi  \equiv \bigg( \bigoplus_\ell \Hi_\ell \bigg) \oplus \Hi_R = \bigg( \bigoplus_\ell \Hi_{\ell}^{(A)} \otimes \Hi_{\ell}^{(B)} \bigg) \oplus \Hi_R,  
\eeq 
in such a way that
\begin{equation}
\Ai=\bigg( \bigoplus_\ell \mathcal{B}(\Hi_{\ell}^{(A)})\otimes {\mathbb I}_{\ell}^{(B)}\bigg) \oplus {\mathbb O}_R,
\label{algdec}
\end{equation}
where ${\mathbb I}_{\ell}^{(B)}$ represents the identity operator on the factor 
$\Hi_{\ell}^{(B)}$ and ${\mathbb O}_R$ the zero operator on $\Hi_R$, respectively. Relative to the same decomposition, the {\em commutant} $\Ai'$ of $\Ai$ in $\BH$, given by $\Ai'\equiv \{  Y\,|\, [Y,X]=0,\: \forall X \in \Ai  \}$, has the dual structure
\begin{equation}
\Ai' =\bigg( \bigoplus_\ell {\mathbb I}_{\ell}^{(A)} \otimes \mathcal{B}(\Hi_{\ell}^{(B)}) \bigg) \oplus \mathcal{B}(\hilbert_R).
\end{equation}

Consider now a density operator $\rho \in {\mathcal D}(\Hi)$ {such that $\supp({\cal A})\subseteq\supp(\rho)$, 
where for a generic operator space $W$ the support is henceforth defined as $\supp(W) \equiv \cup_{O\in W} 
\supp(O)$. }
It then follows that 
\begin{equation}
\label{distorted}
\Ai_\rho \equiv \rho^{\frac{1}{2}} \, \Ai \,\rho^{\frac{1}{2}} = 
\{ Y \,|\, Y=\rho^{\frac{1}{2}}X\rho^{\frac{1}{2}}, \:X \in \Ai\} \subseteq \mathcal{B}(\hilbert) 
\end{equation}
is a $\dag$-closed subalgebra with respect to the modified operator product $X_1\circ_\rho X_2 \equiv X_1\rho^{-1} X_2,$ 
where the inverse is in the sense of Moore-Penrose \cite{horn-johnson} if $\rho$ is not invertible.
In the QIP literature, an associative algebra like in Eq. \eqref{distorted}, which may be thought of as arising from 
a standard associative algebra $\Ai$ upon replacing each identity factor in Eq. \eqref{algdec} with a fixed matrix 
$\tau_{\ell}^{(B)}$ in each factor has been termed a {\em distorted algebra} \cite{viola-IPS,viola-IPSlong}.
In particular, we shall call $\Ai_\rho$ a {\em $\rho$-distorted algebra}, and refer to the map 
$\Phi_\rho(X) \equiv \rho^{\frac{1}{2}}X\rho^{\frac{1}{2}}$ as a ``distortion map''. The $\rho$-distorted algebra 
generated by a set of operators $X_1,\ldots, X_k$ will be correspondingly denoted by 
$\Ai_\rho\equiv \alg_\rho (\{X_1,\ldots , X_k\})$.

\subsection{Fixed points of quantum dynamical semigroups}
\label{sub:fixed}

States that are invariant (aka stationary or ``fixed'') under the dissipative dynamics of interest will 
play a central role in our analysis. Let $\fix(\Ti)$ indicate the set of fixed points of a CP map $\Ti$; 
when $\Ti_t =e^{\Li t}$ for $t\geq 0$, then clearly $\fix(\Ti_t)=\ker(\Li)$. In this section, {we 
summarize relevant results on the structure of fixed-point sets for CPTP maps, and slightly extend them to 
continuous-time QDS evolutions.} 

Recall that fixed points of {\em unital} CPTP maps form a $\dag$-closed algebra: this stems from 
the fact that $\alg\{\Ti\}=\alg\{\Ti^\dag\}$, along with the following result (see e.g. Theorem 6.12 in \cite{wolf-notes}):

\begin{lemma}
\label{thm:commutantkernelkraus}
Given a CPTP map $\Ti,$ the commutant $\alg\{\Ti\}'$ is contained in  $\fix(\Ti^\dag)$. In particular, if there exists a  
{positive-definite state} $\rho>0$ in $\fix(\Ti)$, then
\begin{equation}
\alg\{\Ti\}'=\fix(\Ti^\dag).
\end{equation}
\end{lemma}

\noindent 
If ${\cal T}$ is CPTP and unital, its dual map always admits the identity as a fixed point of full rank. It then follows that 
$\fix(\Ti)=\alg\{\Ti\}'$ \cite{kribs2003,viola-IPSlong}. A similar result can be established for QDS generators (Theorem 7.2, \cite{wolf-notes}):

\begin{lemma}
\label{thm:commutantkernel}
Given a QDS generator $\lind,$ the commutant $\alg\{\lind\}'$ is contained in the kernel of $\lind^\dag$. In particular, if $\lind(\rho)=0$ for some $\rho>0$, then
\begin{equation}
\textup{alg}\{\lind\}'=\ker(\lind^\dag).
\end{equation}
\end{lemma}

A key result to our aim is that, in general, the set of fixed points of a QDS has the structure of a distorted algebra. 
The following characterization is known for arbitrary (non-unital) CPTP maps 
(see e.g. Corollary 6.7 in \cite{wolf-notes}):

\begin{thm}
\label{thm:dualkernelkraus}
\label{thm:generalfixedpointkraus}
Given a CPTP map $\Ti$ and a full-rank fixed point $\rho$,
\begin{equation}
\fix(\Ti)=\rho^\frac{1}{2}\,\fix(\Ti^\dag)\,\rho^\frac{1}{2},
\end{equation}
Moreover, with respect to the decomposition $\fix(\Ti^\dag)=
\bigoplus_\ell {\cal B}(\Hi_{\ell}^{(A)}) \otimes {\mathbb I}_{\ell}^{(B)}$, 
we have
\beq\label{rhostr}\rho = \bigoplus_\ell \rho_{\ell}^{(A)} \otimes \tau_{\ell}^{(B)},\eeq
where $\rho_{\ell}^{(A)}$ and $\tau_{\ell}^{(B)}$ are {\em full-rank} density operators of appropriate dimension.
\end{thm}

\noindent 
Building on the previous results, an analogous statement can be proved for QDS dynamics:

\begin{thm} 
{\bf (QDS fixed-point sets, full-rank case)}  
\label{thm:dualkernel}
Given a QDS generator $\lind$ and a full-rank fixed point $\rho$,
\begin{equation}
\ker(\lind)=\rho^\frac{1}{2}\, \ker(\lind^\dag)\, \rho^ \frac{1}{2}.
\end{equation} 
Moreover, with respect to the decomposition 
$\ker(\Li^\dag)=\bigoplus_\ell {\cal B}(\Hi_{\ell}^{(A)}) \otimes {\mathbb I}_{\ell}^{(B)}$, 
we have 
\beq\label{rhostr1}\rho  = \bigoplus_\ell \rho_{\ell}^{(A)} \otimes \tau_{\ell}^{(B)},\eeq
\noindent 
where $\rho_{\ell}^{(A)}$ and $\tau_{\ell}^{(B)}$ are {\em full-rank} density operators of appropriate 
dimension.
\end{thm}

\vspace*{2mm}
\noindent
{\bf Proof:}
In order for $\{e^{{\cal L}t}\}_{t\geq 0}$ to be a QDS, and thus a semigroup of trace-norm contractions \cite{alicki-lendi}, 
$\Li$ must have spectrum in the closed left-half of the complex plane and no purely imaginary eigenvalues with multiplicity. 
It is then easy to show, by resorting to its Jordan decomposition \cite{wolf-notes}, that the following limit exists:
\[\Ti_{\infty} \equiv \lim_{t\rightarrow\infty}\frac{1}{t}\int_0^t e^{\Li \tau} d\tau.\]
Being {the limit of convex combination of CPTP maps, which form a closed convex set, $\Ti_{\infty}$ it also CPTP.} 
Furthermore, $\Ti_\infty$ projects onto $\ker(\Li),$ namely, $\fix(\Ti_\infty)=\ker(\Li)$, and $\Ti_\infty$ 
has only eigenvalues $0,1$ with simple Jordan blocks. {Similarly, it follows that the unital CP map
$ \Ti^\dag_{\infty}  \equiv (\Ti_{\infty})^\dag$}
projects onto $\ker({\Li^\dag}).$
Using these facts along with Theorem \ref{thm:dualkernelkraus}, we then have:
\[\ker(\lind)=\fix(\Ti_{\infty})=\rho^\frac{1}{2}\,\fix(\Ti_\infty^\dag)\,\rho^\frac{1}{2}=\rho^\frac{1}{2}\,
\ker(\Li^\dag)\, \rho^\frac{1}{2}. \]
{The structure of the fixed point, Eq. \eqref{rhostr1}, follows from 
Theorem \ref{thm:dualkernelkraus} applied to $\Ti_\infty.$}
\qed

\vspace*{1mm}

{The above two theorems make it clear that, given discrete- or continuous-time CPTP dynamics 
admitting a full-rank invariant state $\rho$, the fixed-point sets $\fix(\Ti)$ and $\ker({\Li})$ are a 
$\rho$-distorted algebra with structure
\beq
\label{CPTPfixedpoints}
{\cal A}_\rho=\bigoplus_\ell {\cal B}(\Hi_{\ell}^{(A)}) \otimes \tau_{\ell}^{(B)}, \eeq
where the states $\tau_{\ell}^{(B)}$ are the same for every element in $\fix(\Ti)$ or $\ker({\Li})$.
In addition, since $\rho$ has a {\em compatible} block structure [Eq. \eqref{rhostr1}], 
it is immediate to see that $\fix(\Ti)$ and $\ker({\Li})$ are invariant with respect to the action of the linear map 
${\cal M}_{\lambda}(X) \equiv  \rho^{\lambda} X \rho^{-\lambda}$ for any $\lambda \in \mathbb{C}$, and 
in particular for the {\em modular group} $\{{\cal M}_{i\phi}\}$ \cite{petz}. The same holds for the fixed points of the dual dynamics. 

In fact, we can show that modular invariance 
is also a sufficient condition for a distorted algebra to be a fixed-point set for a CPTP map that 
fixes $\rho$, as relevant to the problem of designing stabilizing dynamics for $\rho$.
In order to do this, we need a result by Takesaki \cite{takesaki}, which we give 
in its finite-dimensional formulation (adapted from \cite{petz}, Theorem 9.2):

\begin{thm}
\label{thm:takesaki}
Let ${\cal A}$ be a $\dag$-closed subalgebra of ${\cal B}(\Hi),$ and $\rho$ a full-rank density operator. 
Then the following are equivalent: \\
(i) There exists a unital CP map $\Ei^\dag$ such that  $\fix(\Ei^\dag)={\cal A},$ 
$(\Ei^\dag)^2=\Ei^\dag$ and $\Ei(\rho)=\rho.$ \\
{(ii) ${\cal A}$ is invariant with respect to ${\cal M}_{\frac{1}{2}}$, that is, for every $X\in{\cal A},$
$\rho^{\frac{1}{2}}X\rho^{-\frac{1}{2}}\in{\cal A}. $}
\end{thm}

\vspace*{1mm}

{These conditions are equivalent to saying that the map $\Ei^\dag$ is a {\em conditional expectation} on $\Ai$ 
that preserves $\rho$. We can then prove the following:

\begin{thm}
\label{thm:modinvariance}
{\bf (Existence of $\rho$-preserving dynamics)}
Let $\rho$ be a full-rank density operator. A distorted algebra ${\cal A}_\rho$ admits a CPTP map $\Ti$ such that $\fix(\Ti)={\cal A}_\rho$ if and only if it is invariant for ${\cal M}_{\frac{1}{2}}.$
\end{thm}
{\bf Proof.}
First, notice that, if $\Ai_\rho$ is a distorted algebra, then it is  invariant for ${\cal M}_{\frac{1}{2}}$ if and only if the ``undistorted'' algebra
 $\Ai \equiv \rho^{-\frac{1}{2}}\Ai_\rho\rho^{-\frac{1}{2}}$ is invariant for ${\cal M}_{\frac{1}{2}}$. 
This follows from the fact that ${\cal M}_{\frac{1}{2}}$  
commutes with both the distortion map and its inverse. In particular, if $\Ai_\rho$ is invariant for ${\cal M}_{\frac{1}{2}}$, we have:
\[{\cal M}_{\frac{1}{2}}(\Ai)={\cal M}_{\frac{1}{2}}(\rho^{-\frac{1}{2}}\Ai_\rho\rho^{-\frac{1}{2}})=\rho^{-\frac{1}{2}}{\cal M}_{\frac{1}{2}}(\Ai_\rho)\rho^{-\frac{1}{2}}\subseteq \rho^{-\frac{1}{2}}\Ai_\rho\rho^{-\frac{1}{2}}=\Ai.\]
Thus, by Theorem \ref{thm:takesaki}, a unital CP projection $\Ei^\dag$ onto $\Ai$ exists whose adjoint preserves 
$\rho$. By Theorem \ref{thm:dualkernelkraus}, the CPTP dual $\Ei \equiv \Ti$ is such that $\fix{(\Ti)}=\Ai_\rho$, as
desired.

To prove the other implication, it is sufficient to notice that Eq. \eqref{rhostr} implies that ${\cal M}_{\frac{1}{2}}$ 
leaves $\Ai=\fix(\Ti^\dag)$ invariant, and thus
\[ \hspace{2cm}
{\cal M}_{\frac{1}{2}}(\Ai_\rho)={\cal M}_{\frac{1}{2}}(\rho^{\frac{1}{2}}\Ai_\rho\rho^{\frac{1}{2}})=\rho^{\frac{1}{2}}{\cal M}_{\frac{1}{2}}(\Ai)\rho^{\frac{1}{2}}\subseteq \rho^{\frac{1}{2}}\Ai\rho^{\frac{1}{2}}=\Ai_\rho.
\hspace{3cm}\]}}\qed 
If the dynamics admit no full-rank fixed state, we may 
restrict to the support of a given fixed point, which is an invariant subspace for the Schr\"odinger's-picture evolution: 

\begin{thm}
\label{thm:dualkernel-general} 
{{\bf (QDS fixed-point sets, general case)}  
Given a finite-dimensional QDS generator $\lind,$ 
and a maximal-rank fixed point $\rho$ with $\tilde\Hi\equiv \supp(\rho)$, let 
$\tilde\Li$ denote the reduction of $\Li$ to $\mathcal{B}(\tilde\Hi)$.  We then have }
\begin{equation}
\ker(\lind)=\rho^\frac{1}{2}\,(\ker(\tilde\lind^\dag)\oplus {\mathbb O}) \,\rho^\frac{1}{2}.
\end{equation}
\end{thm}

\vspace*{10pt}
\noindent
{\bf Proof:}
{For any $\rho\in\ker(\Li)$, the subspace $\tilde\Hi \equiv \supp(\rho)$ 
is invariant for the dynamics \cite{ticozzi-markovian}.} 
Assume that $\Li=\Li(H,\{L_k\})$ and let $\tilde\Pi:\Hi\rightarrow\tilde\Hi$ 
denote the partial isometry onto $\tilde\Hi.$ Define the 
reduced (projected) operators $\tilde \rho \equiv \tilde\Pi \rho \tilde\Pi^\dag,$ 
$\tilde H \equiv \tilde\Pi H \tilde\Pi^\dag$, and $\tilde L_k \equiv \tilde\Pi L_k \tilde\Pi^\dag.$ 
The dynamics inside $\tilde\Hi$ is then determined by the corresponding projected Liouvillian 
$\tilde\Li(\tilde H, \{\tilde L_k\})$ \cite{ticozzi-QDS}, and $\tilde\rho$ is, by construction, a full-rank state
for this dynamics. Hence, the fixed-point set $\ker(\tilde\Li)$ is the distorted algebra 
{$\ker(\tilde\Li)=\rho^{\frac{1}{2}} \ker(\tilde\Li^\dag)\rho^{\frac{1}{2}}.$ }

Consider now a {maximal-rank} fixed point, satisfying {$\supp(\rho)=\tilde\Hi= \supp(\ker(\Li))$.
It then follows from Theorem 9 in \cite{ticozzi-markovian} that $\tilde\Hi$ 
is not only invariant but also attractive for the dynamics}. This means that 
$$ \lim_{t\rightarrow\infty} \Tr \Big(\tilde{\Pi}^\perp e^{\Li t} (\rho_0) \Big)=0 , \quad \forall \rho_0 \in {\mathfrak D}(\Hi).$$
\noindent
With $\tilde\Hi$ being attractive, we have that $\ker(\Li)$ can have support {\em only} in $\tilde\Hi,$ and 
can thus be constructed by appending the zero operator on $\tilde\Hi^\perp,$ so that, using Theorem \ref{thm:dualkernel}:
\[\hspace{3.6cm}
\ker(\Li)=\ker(\tilde\Li)\oplus {\mathbb O} =\rho^\frac{1}{2}\,(\ker(\tilde\Li^\dag)\oplus {\mathbb O})\; \rho^\frac{1}{2}.
\hspace{4cm}\mbox{} \]\qed

\noindent 
In the above proof, we made the construction explicit in terms of a representation  $\Li=\Li(H,\{L_k\})$ in order to make it clear that the result does not hold if we consider $\rho^\frac{1}{2}\,\ker(\Li^\dag)\,\rho^\frac{1}{2},$ since $\tilde\Hi$ need {\em not} be invariant for $\Li^\dag.$ {Again, it follows that $\rho$ admits a block decomposition as in Eq. \eqref{rhostr1}, compatible with that of 
{$\ker({\tilde{\Li}^\dag})$} on its support.}

\subsection{Quasi-local semigroup dynamics}
\label{sub:ql} 

Throughout this paper, the open quantum system of interest will consist of a finite number $n$ of distinguishable 
subsystems, defined on a tensor-product Hilbert space
$$\Hi=\bigotimes_{a=1}^n\Hi_a, \quad \text{dim}(\Hi_a)=d_a, \, \text{dim}(\Hi)=D. $$
\noindent 
As in \cite{ticozzi-ql,ticozzi-steady}, we shall introduce quasi-locality constraints on the system's evolution 
by specifying a list of {\em neighborhoods}, namely, groups of subsystems on which operators are allowed to 
``act simultaneously''. Mathematically, neighborhoods $\{ {\cal N}_j \}$ may be specified as subsets of the set of 
indexes labeling the subsystems, that is,
\[{\cal N}_j\subseteq\{1,\ldots,n\}, \quad j=1,\ldots, M.\]
Each neighborhood induces a bipartite tensor-product structure of $\Hi$ as
\begin{equation}
\label{bipart}
\Hi = \Hi_{{\cal N}_j} \otimes \Hi_{\overline{\cal N}_j}, \quad \quad 
\Hi_{{\cal N}_j} \equiv \bigotimes_{a\in{\cal N}_j}\Hi_a, 
\quad \Hi_{\overline{\cal N}_j} \equiv \bigotimes_{a\notin{\cal N}_j}\Hi_a.
\end{equation}
Likewise, with a neighborhood structure $\neigh \equiv \{{\cal N}_j\}$ in place, any state 
$\rho\in\DH$ uniquely determines a list of {\em reduced neighborhood states} $\{ \rho_{{\cal N}_j} \}$:
\begin{equation} 
\label{redstate} 
\rho_{{\cal N}_j} \equiv \mbox{Tr}_{\overline{\cal N}_j}(\rho),
\quad \rho_{{\cal N}_j} \in {\mathfrak D}(\Hi_{{\cal N}_j}), \;\; j=1,\ldots, M,
\end{equation}
where $\Tr_{\overline{\cal N}_j}$ indicates the partial trace over $\Hi_{\overline{\cal N}_j}$.
Quasi-local dynamical constraints may be specified by requiring compatibility 
with the bipartitions in \eqref{bipart}, in the following sense:

\begin{defin} {\bf (Neighborhood operator)}
An operator $X\in \BH$ is a {\em neighborhood operator} relative to a given neighborhood structure 
$\neigh$ if there exists $j$ such that the action of $X$ is non-trivial only on $\Hi_{{\cal N}_j}$, that is:
\[ X = X_{{\cal N}_j}\otimes {\mathbb I}_{\overline{\cal N}_j}, \] 
where ${\mathbb I}_{\overline{\cal N}_j}$ is the identity operator on $\Hi_{\overline{\cal N}_j}$.
\end{defin} 

\noindent 
A similar definition may be given for neighborhood CPTP maps and generators. 
The relevant quasi-locality notion for QDS dynamics is then the following:

\begin{defin} 
\label{def:qll}
{\bf (QL semigroup)}
A QDS generator $\Li$ is {\em Quasi-Local (QL)} relative to a given neighborhood structure $\neigh$ if 
it may be expressed as a sum of neighborhood generators:
\end{defin}
\begin{equation}
\Li = \sum_j \Li_j , \quad \Li_j \equiv \Li_{{\cal N}_j} \otimes {\cal I}_{\overline{\cal N}_j}. 
\label{QLL}
\end{equation}

\noindent 
Quasi-locality of a Liouvillian is well-defined, as the structural property in Eq. (\ref{QLL}) is 
defined independently of a particular representation of the generator.
In terms of an explicit representation, the above definition is equivalent to 
requiring that there exists some choice $\Li \equiv \Li( H, \{L_k\})$, such that
each Lindblad operator $L_k$ is a neighborhood operator and the Hamiltonian may be 
expressed as a sum of neighborhood Hamiltonians, namely:
\[ L_k = L_{k, \neigh_j} \otimes {\mathbb I}_{\overline{\cal N}_j},\quad 
H = \sum_j H_j, \quad H_j \equiv H_{{\cal N}_j}\otimes I_{\overline{\cal N}_j}.\]
\noindent
A Hamiltonian $H$ of the above form is called a {\em QL Hamiltonian} (often ``few-body,'' in the 
physics literature)\footnote{In particular, the notions of 
neighborhood Hamiltonian and QL Hamiltonian reduce to the standard uni-local and local ones 
for non-overlapping neighborhoods, see e.g. \cite{Barnum2003}.}.
Mathematically, this denomination is natural given that, for the limiting case of closed-system 
dynamics, a QL Hamiltonian so defined automatically induces a QL (Lie-)group action consistent with 
Eq. (\ref{QLL}), with $\Li_j \equiv i \,{\text{ad}_{H_j} (\cdot)} = i \,[H_j, \cdot]$.

\vspace*{1mm}

\noindent 
{\bf Remark 1.} 
The above QL notion is appropriate to describe any locality constraint that may be associated with a 
spatial lattice geometry and {\em finite interactions range} (e.g., spins living on the vertices 
of a graph, subject to nearest-neighbor couplings). 
QL semigroup dynamics have also been considered under less restrictive 
assumptions on the spatial decay of interactions \cite{Cubitt2015}, and yet different QL notions may be 
potentially envisioned (e.g. based on locality in ``momentum space'' or relative to ``error weight''). 
The present choice provides the simplest physically relevant setting that allows for a direct 
linear-algebraic analysis.  
We stress that, due to the freedom in the representation of the QDS generator, QL semigroup dynamics 
may still be induced by Lindblad operators that are not manifestly of neighborhood form. In principle, it 
is always possible to check the QL property by verifying whether a QDS generator ${\cal L}$ has 
components only in the (super-)operator subspace spanned by QL generators. 
While it may be interesting to determine more operational and efficient QL criteria in 
specific cases, in most practical scenarios (e.g. open-system simulators \cite{barreiro2}) available Lindblad operators 
are typically specified in a preferred neighborhood form from the outset.

\subsection{Quasi-local stabilizability: Prior pure-state results and frustration-free semigroup dynamics}
\label{subsec:qlsdefns}

Our main focus will be on determining conditions under which a certain state of interest, 
$\rho$, is guaranteed to be invariant and the unique asymptotically stable state for some QL dynamics. 
Formally, an invariant state $\rho\in{\mathfrak D}(\Hi)$ for a QDS with generator $\Li$ is said to be 
{\em Globally Asymptotically Stable} (GAS) if 
\begin{equation}
\lim_{t\rightarrow +\infty} e^{{\cal L}t}(\rho_0)= \rho, \quad \forall \rho_0 \in {\mathfrak D}(\Hi).
\label{eq:gas}
\end{equation}
{For QDS dynamics not subject to QL constraints
it is known that a state is GAS if and only if it is the {\em unique} fixed point 
\cite{wolf-notes,ticozzi-generic}. 
A definition of stabilizable states relevant to our constrained setting may be given as follows:}

\begin{defin} 
{\bf (QLS state)} 
{A state $\rho\in {\mathfrak D}(\Hi)$ is {\em Quasi-Locally Stabilizable (QLS)} relative to a neighborhood 
structure $\neigh$ if there exists a QL generator ${\cal L}$ for which $\rho$ is GAS.}
\end{defin}

Existing work has so far focused on stabilizability of a {\em pure state}, with special emphasis on steady-state entanglement \cite{ticozzi-ql,Kraus2008,ticozzi-steady}. While even in this case, in general, a careful balancing of Hamiltonian and dissipative action is essential, a simple
 yet non-trivial stabilization setting arises by further requiring that the target can be made QLS by a generator without a Hamiltonian component, namely, by using dissipation alone. Given the freedom in the representation of a QDS generator, in order to formalize this additional constraint we introduced a {\em standard representation} for a generator $\Li(H,\{ L_k\})$ that fixes a pure state $\rho\equiv \ket{\Psi}\bra{\Psi}\in{\mathfrak D}(\Hi),$ as in the following result (Corollary 1 in \cite{ticozzi-steady}):

\begin{Prop}
\label{standardform} 
If a generator ${\cal L}(H, \{ L_k\})$ makes $\rho=\ket{\Psi}\bra{\Psi}$ GAS, then the same
generator can be represented in a standard form ${\cal L}(\tilde{H},\{ \tilde{L}_k \})$, in such a way that
$\tilde{H}\ket{\Psi}=h\ket{\Psi},$ $h\in\mathbb{R}$ and $\tilde{L}_k\ket{\Psi}=0$, for all $k$.
\end{Prop} 

\noindent 
In the standard representation, the target $\ket{\Psi}\in \ker(\tilde{L}_k)$ may thus be seen as a common 
``dark state'' for all the noise operators, borrowing from quantum-optics terminology. 
With this in mind, a pure state $\rho=\ket{\Psi}\bra{\Psi}$ may be defined as {\em Dissipatively Quasi-Locally 
Stabilizable} (DQLS) if it is QLS with $\tilde{H}\equiv 0$ and QL noise operators $\{ \tilde{L}_k \}$ 
in standard form. Notice that such a definition implies that $\rho$ is invariant for the dynamics relative to {\em each} neighborhood, namely, $\rho\in\ker(\Li( \{\tilde L_k \})),$ for each $k.$ Building on this QL-invariance condition allows for proving the following characterization of DQLS states \cite{ticozzi-ql}:

\begin{thm}
\label{mainthm} 
A pure state $\rho=|\Psi\rangle\langle \Psi| \in{\mathfrak D}(\Hi)$ is DQLS relative to $\Ni$ if and only if 
\begin{equation}
\label{GAScond} 
\supp(\rho) = \bigcap_k \supp(\rho_{\neigh_k}\otimes \mathbb{I}_{\overline{\neigh}_k}).
\end{equation}
\end{thm}

\noindent
{\bf Remark 2.} The proof of the above result includes the construction of a set of stabilizing Lindblad operators 
$\tilde L_k$ that make $\rho$ DQLS, also implying that {\em one} such operator per neighborhood always suffices. 
It is easy to show that any rescaled version of the same operators, $\tilde L_k' \equiv r_k \tilde L_k$, also 
yield a stabilizing QL generator $\Li'=\sum_k |r_k|^2 \Li_k \equiv \sum_k \gamma_k \Li_k$ {-- incorporating
``model  ($\gamma$-)robustness,'' in the terminology of \cite{ticozzi-QDS}.
\label{gamma}

\vspace*{1mm}

However, the reasoning followed for QL stabilization of a pure state does {\em not} extend naturally to a general, 
mixed target state.} The main reason is that the standard form, hence the DQLS 
definition itself, do {\em not} have a consistent analogue for mixed states. A major simplification if $\rho$ is pure 
stems from the fact that it is straightforward to check for invariance, directly in terms of the generator components 
(see Proposition 1 in \cite{ticozzi-steady}); for general $\rho$, we seek a definition that extends the DQLS 
notion, and that similarly allows for explicitly studying what the invariance of $\rho$ means at a QL level.
A natural choice is to restrict to the class of {\em frustration-free dynamics}. That is, in addition to the QL 
constraint, we demand that {\em each} QL term in the generator leave the state of interest invariant. 
Formally, we define \cite{Brandao2014}:

\begin{defin} {\bf (FF generator)} 
{A QL generator $\mathcal{L}=\sum_{j}\mathcal{L}_{j}
$ is \emph{Frustration Free} (FF) relative to a neighborhood structure $\Ni =\{ \Ni_j\}$ if any invariant state 
$\rho\in\ker (\Li)$ also satisfies neighborhood-wise invariance, namely, $\rho \in \ker(\Li_j)$ for all $j$.}
\end{defin}

\noindent 
Beside allowing for considerable simplification, FF dynamics are of practical interest because 
they are, similar to the DQLS setting, robust to certain perturbations. 
As in Remark \ref{gamma}, given a QL generator $\mathcal{L}=\sum_{j}\mathcal{L}_{j}$, define 
a ``neighborhood-perturbed'' QL generator $\mathcal{L}'=\sum_{j}\lambda_{j}\mathcal{L}_{j}$, with 
$\lambda_{j}\in\mathbb{R}^+$. If $\mathcal{L}$ is FF, then $\mathcal{L}(\rho)=0$ implies $\mathcal{L}_{j}(\rho)=0$ 
for each $j$; therefore, $\lambda_{j}\mathcal{L}_{j}(\rho)=0$ for each $j$, and thus $\mathcal{L}'(\rho)=0$ as well. 
{Were $\mathcal{L}$ not FF, then the kernel of $\mathcal{L}$ would not be robust 
against such neighborhood-perturbations in general. With these motivations, we introduce the notion 
of QL stability that we analyze for the remainder of this paper:}

\begin{defin}{\bf (FFQLS state)}
{A state $\rho\in{\mathfrak D}(\Hi)$ is {\em Frustration-Free Quasi-Locally Stabilizable} (FFQLS) relative to 
a neighborhood structure $\neigh$ if it is QLS with a stabilizing generator ${\cal L}$ that is FF.}
\end{defin}

\noindent 
Remarkably, studying FFQLS states will allow us to recover the results for DQLS pure states as a special case. 
In fact, {\em a pure state is FFQLS if and only if it is DQLS}. Since the proof of this claim relies on 
results that we develop in the following sections, it is presented in Appendix A. 
To summarize, the above definition consists of four distinct mathematical conditions that the generator $\Li$ 
of the dissipative dynamics must obey for a given target invariant state $\rho$:
\begin{itemize}
\item(QDS): $\lind$ is a generator of a CPTP continuous semigroup;
\item(QL): $\Li$ is QL, that is, 
$\lind=\sum_j\lind_j,$ with $\lind_j={\lind}_{\neigh_j}\otimes\mathcal{I}_{\overline{\neigh}_j}$;
\item(GAS): $\rho$ is GAS, or equivalently $\ker(\lind)=\textup{span}(\rho)$; 
\item(FF): $\Li$ is FF, namely $\ker(\lind_j)\geq\textup{span}(\rho)$ for all $\neigh_j$.
\end{itemize}
The problem we are interested in is to determine necessary and sufficient conditions for a given state to be 
FFQLS and, if so, to design QL FF dynamics that achieves the task.

\section{Frustration-Free Stabilizable States: Necessary Conditions} 
\label{sec:necessary}

In this section, we derive necessary conditions for a target state to be FFQLS. Frustration-freeness requires such a state to be in the kernel of each neighborhood generator. We show that, if a neighborhood generator is to leave a {\em global}  state invariant, the size and structure of its kernel are constrained; in general, the kernel will be larger than the span of the {\em reduced} neighborhood state (as a vector in Hilbert-Schmidt space). However, if the target state is to be the {\em unique} fixed point of the QDS dynamics, then the intersection of all the neighborhood-generator kernels must coincide with the span of the target state. We shall show in Section \ref{subsec:fullrank} that this condition is also sufficient for a generic (full-rank) state to be FFQLS.

\subsection{Linear-algebraic tools}\label{sub:linear}

Recall that, given a tensor product of two inner-product spaces $V=V_A\otimes V_B$ and a vector $v\in V$, 
a \emph{Schmidt decomposition} of $v$ is any decomposition
$$v=\sum_i \lambda_i a_i\otimes b_i , $$
where $a_i\in V_A$, $b_i\in V_B$, $\lambda_i>0$, and $\{a_i\}$, $\{b_i\}$ are each orthonormal sets of vectors.
There are two instances of Schmidt decomposition which are relevant in our context, both well known within QIP 
\cite{Nielsen2003}. The first is the Schmidt decomposition of a bipartite pure state $\ket{\psi}\in\hilbert_{A}\otimes\hilbert_{B}$, namely, 
$$ \ket{\psi}=\sum_i\lambda_i\ket{a_i}\otimes\ket{b_i}. $$ 
The second is the so-called operator-Schmidt decomposition, whereby a bipartite operator $M\in\mathcal{B}(\hilbert_A\otimes\hilbert_B)=\mathcal{B}(\hilbert_A)\otimes\mathcal{B}(\hilbert_B)$ is factorized in terms of 
elements in the vector spaces $\mathcal{B}(\hilbert_A)$ and $\mathcal{B}(\hilbert_B)$, relative to the 
Hilbert-Schmidt inner product. Specifically, 
$$  M=\sum_i \lambda_i A_i\otimes B_i, $$
where $A_i\in \mathcal{B}(\hilbert_A)$, $B_i\in \mathcal{B}(\hilbert_B)$, $\lambda_i>0$, and 
$\mbox{Tr}(A^{\dagger}_i A_j)=\mbox{Tr}(B^{\dagger}_i B_j)=\delta_{ij}$.

Building on the concept of Schmidt decomposition, we introduce the Schmidt span:

\begin{defin} {\bf (Schmidt span)} 
{Given a tensor product of two inner product spaces $V=V_A\otimes V_B$ and a vector $v\in V$ with Schmidt decomposition $v=\sum_i \lambda_i a_i\otimes b_i$, the \emph{Schmidt span} of $v$ relative to $V_A$ is the 
subspace }
\begin{equation}
\Sigma_A(v)=\span \Big \{a_i\in V_A\, |\, v=\sum_ia_i\otimes b_i , \: b_i \in V_B \Big\}. 
\end{equation}
\end{defin}

\noindent
{\blue 
Without referring to a particular tensor-product decomposition, it is possible to show that the Schmidt span is
the image of 
$v$ under partial inner product:
\begin{equation}
\label{eq:partialinnerprod}
\Sigma_A(v) = \{a\in V_A\, |\,  {a}=(\identity_A\otimes {b}^{\dagger}) v \,\,\text{for some}\,\, b\in V_B\}.
\end{equation}
One example is when $V_A\otimes V_B$ is a matrix space, such as 
$\complex^{d_A}\otimes\complex^{d_B\dag}$ (where the latter factor is meant as a space of row vectors). 
In this case, the Schmidt span of a matrix $W\in\complex^{d_A}\otimes\complex^{d_B\dag}$, relative to the first factor 
$\complex^{d_A}$, is simply the \emph{range} of the matrix $W$ (namely, the set of all linear combinations of its column vectors), namely $\Sigma_A(W)=\textup{range}(W)$. Similarly, the Schmidt span of $W$ relative to the second factor 
$\complex^{d_B\dag}$ is the orthogonal complement of the kernel of $W$ or, in other words, the 
\emph{support}: $\Sigma_B(W)=\textup{supp}(W).$
Another example is when $V_A\otimes V_B$ is a bipartite operator space, such as $\mathcal{B}(\hilbert_A)\otimes\mathcal{B}(\hilbert_B)$. The Schmidt span of $M \in \mathcal{B}(\hilbert_A)\otimes\mathcal{B}(\hilbert_B)$ relative to $\mathcal{B}(\hilbert_A)$, i.e. ``on $A$'', is the operator subspace
$\Sigma_A(M)=\{ \trb{B}{(I_A\otimes B) M},\,\, B\in \mathcal{B}(\hilbert_B)\}$.}

The Schmidt span is a useful tool because conditions on how a neighborhood operator is to affect a global state constrains how such an operator must act on the {\em entire} operator-Schmidt span of that state. This intuition is formalized in the following Lemma:

\begin{lemma}
\label{invSS}
{{\bf (Invariance of Schmidt span)}
Given a vector $v\in V_{A}\otimes V_{B}$ and $M_A\in\mathcal{B}(V_A)$, if $(M_A\otimes \identity_B) v=\lambda v$, then $(M_A\otimes\identity_B) v'=\lambda v'$ for all $v'\in\Sigma_{A}(v)\otimes V_B$. In particular:}
\begin{equation}
\textup{span}(v)\leq\ker(M_A\otimes\identity_B)\Rightarrow\Sigma_{A}(v)\otimes V_B\leq \ker(M_A\otimes\identity_B), 
\label{l0}
\end{equation}
and 
\begin{equation}
\textup{span}(v)\leq\fix(M_A\otimes\identity_B)\Rightarrow\Sigma_{A}(v)\otimes V_B\leq \fix(M_A\otimes\identity_B). 
\label{l1}
\end{equation}
\label{thm:localkernel}
\end{lemma}
\noindent
{\bf Proof:}
Consider the Schmidt decomposition of $v$,  $ v=\sum_i \gamma_i a_i\otimes b_i,$
where $a_i\in V_A$, $b_i\in V_B$, $\gamma_i>0$, and $\{a_i\}$, $\{b_i\}$ are each orthonormal sets of vectors. 
Applying the eigenvalue equation for $M_A$ to this yields 
$$ \sum_i \gamma_i M_A a_i\otimes b_i = \lambda \sum_i \gamma_i a_i\otimes b_i.$$
\noindent 
Multiplying both sides by $\identity_A\otimes b_j^{\dagger}$, where $b_j^{\dagger}$ is the dual vector of $b_j$, 
selects out the $V_A$-factor of the $i$th term, i.e., $M_A a_i = \lambda a_i$.
This holds for each $i$ and any linear combination of the $a_i$s. By definition, the Schmidt span of $v$ is 
$\Sigma_A(v)=\textup{span}\{a_i\}$. Denoting $\{\beta_i\}$ a basis for $V_B$, we may write any 
$v'\in\Sigma_A(v)\otimes V_B$ as $v'=\sum_{ij}\mu_{ij}a_i\otimes\beta_j$.
Applying $(M_A\otimes\identity_B)$ to this we obtain
$$ (M_A\otimes\identity_B)v'=\sum_{ij}\mu_{ij}M_Aa_i\otimes\beta_j=\lambda v'.$$
\noindent 
Thus, all elements in $\Sigma_A(v)\otimes V_B$ have eigenvalue $\lambda$ with 
respect to $M_A\otimes\identity_B$, as claimed. Eqs. (\ref{l0}) and (\ref{l1}) follow by specializing the 
above result to $\lambda= 0$ and $\lambda =1$, respectively.
\qed

\subsection{Invariance conditions for quasi-local generators}
\label{sub:invariance}

As remarked, we require the global dynamics to be FF. This simplifies considerably the analysis, as global invariance of the target 
state is possible {\em only if} the latter is invariant for each  neighborhood generator. Therefore, we examine  the properties of 
a neighborhood generator that ensure the target state $\rho$ to be in its kernel. Note that if $\rho$ is factorizable 
relative to the neighborhood structure (i.e., a pure or mixed product state), $\rho$ is invariant as long as each 
factor of $\rho$ if fixed. Each such reduced neighborhood state can then be made not only invariant but also 
attractive by a neighborhood generator, if the reduced states are the {\em only} elements
in the kernels of the corresponding ${\cal L}_{\neigh_j}$. This automatically makes the global factorized 
state also GAS. In other words, if $\rho$ is factorizable, then QL stabilizability is guaranteed.
If the target state is non-factorizable (in particular, entangled), the above scheme need not work; 
a non-factorizable state will have some operator Schmidt spans with dimension greater than one. 
The following Corollary, which follows from Lemma \ref{thm:localkernel}, illustrates 
the implication of quasi-locally fixing a state with non-trivial operator-Schmidt spans: 

\begin{cor}
Let $\lind_{j}={\lind}_{\neigh_j}\otimes\mathcal{I}_{\overline{\neigh}_j}$ be a neighborhood Liouvillian. 
If $\rho\in\ker(\Li_j),$ then it must also be that 
\[\Sigma_{\neigh_j}(\rho)\otimes\mathcal{B}(\hilbert_{\overline{\neigh}_j})\leq\ker(\Li_j).\] 
\end{cor}

\noindent
Accordingly, if each neighborhood generator $\lind_{j}$ is to fix a non-factorizable $\rho$ (as is necessary for global invariance with FF dynamics), then each neighborhood generator must be constructed to leave invariant, in general, a larger space of operators -- specifically, the corresponding neighborhood operator-Schmidt span of $\rho$. 

However, leaving {\em only} the Schmidt spans invariant is, in general, {\em not} possible if the dynamics 
are to be CPTP, since a Schmidt span {need not be a distorted algebra} (as required by Theorem 
\ref{thm:dualkernel-general}). We show that, in order for $\rho$ to be in the kernel of a valid QL generator, 
it is necessary that the dynamics leave certain {``minimal fixed-point sets'' generated by the Schmidt spans 
invariant as well.  We give the following:

\begin{defin} {\bf (Minimal modular-invariant distorted algebra)} 
{Let $\rho \in {\mathcal D}(\hilbert)$ be a density operator, and $W\subseteq \mathcal{B}(\hilbert)$. 
The {\em minimal modular-invariant distorted algebra generated by $W$} is the smallest $\rho$-distorted algebra 
generated by $W$ which is invariant with respect to ${\cal M}_{\frac{1}{2}} (X) =\rho^{\frac{1}{2}} X \rho^{-\frac{1}{2}}$, 
where the inverse is in the sense of Moore-Penrose if $\rho$ is not full-rank.}
\end{defin}

In the finite-dimensional case that we consider, ${\cal F}_\rho(W)$ can be constructed by the following 
iterative procedure: define ${\cal F}^0 \equiv \alg_\rho(W),$ and compute
\[{\cal F}^{k+1}=\alg_\rho({\cal M}_{\frac{1}{2}}({\cal F}^k)),\]
until ${\cal F}^{k+1}={\cal F}^{k}\equiv {\cal F}_\rho(W).$ 
This particular distorted algebra is the smallest structure whose invariance is required 
if the dynamics are to be CPTP: 

\begin{lemma}
{\bf {(Minimal fixed-point sets)}}
Let $W\leq\mathcal{B}(\hilbert)$ be an operator subspace containing a positive-semidefinite operator 
$\rho$ such that $\textup{supp}(\rho)=\textup{supp}(W).$ If $W \leq \fix(\Ti)$ for a CPTP 
map $\Ti:\mathcal{B}(\hilbert)\rightarrow\mathcal{B}(\hilbert)$, then 
\beq {\cal F}_\rho (W) \leq \fix(\Ti).\eeq
\label{thm:CPTPfix}
\end{lemma}
\noindent
{\bf Proof:} Given the iterative construction of ${\cal F}_\rho(W),$ it suffices to show that if some set $W\subseteq \fix(\Ti)$ includes a density operator with $\textup{supp}(\rho)=\textup{supp}(W),$ then both $\alg_\rho(W)\subseteq \fix(\Ti)$ 
and ${\cal M}_{\frac{1}{2}}(W)\subseteq \fix(\Ti)$, and their support is still equal to $\supp(\rho).$
Since $\Ti$ is a CP linear map, $\fix(\Ti)$ is closed with respect to linear combinations and $\dag$-adjoint. 
We are left to show that $\fix(\Ti)$ is closed with respect to the $\rho$-modified product. 
Consider a partial isometry $V:\textup{supp}(\rho)\rightarrow\hilbert$, and define the reduced map
$$\tilde{\Ti}:\mathcal{B}(\textup{supp}(\rho))\rightarrow\mathcal{B}(\textup{supp}(\rho)), \quad 
\tilde{\Ti}(X)\equiv V^{\dagger}\Ti(VXV^{\dagger})V.$$ 
\noindent
Since $\rho$ is invariant, the set of operators with support contained in $\textup{supp}(\rho)=\textup{supp}(W)$ is an invariant subspace, and thus $\Ti(VV^{\dagger}XVV^{\dagger})=VV^{\dagger}\Ti(VV^\dag{X} VV^{\dagger})VV^{\dagger}.$ By construction, $\tilde{\Ti}$ is CP, TP, and {\em has a full-rank fixed point} $\tilde{\rho}\equiv V^{\dagger}\rho V$.
It follows from Theorems \ref{thm:generalfixedpointkraus} and \ref{thm:modinvariance} that $\fix(\tilde\Ti)$ is a 
$\rho$-distorted algebra; hence, it is closed with respect to the modified product, as well as modular-invariant.
Now, if $X,Y\in W$ are fixed points for $\Ti$, so are $\tilde X = V^\dag XV,\tilde Y = V^\dag Y V^\dag$ for $\tilde \Ti.$ 
Since their adjoint, linear combinations and $\rho$-distorted products are in $\fix(\tilde \Ti)$, we have:
\beqan
 \Ti(X {\rho}^{-1} Y)& =&\Ti(VV^{\dagger}XVV^{\dagger} {\rho}^{-1} VV^{\dagger}YVV^{\dagger})
=VV^{\dagger}\Ti(V\tilde{X} {\tilde{\rho}}^{-1} \tilde{Y}V^{\dagger})VV^{\dagger}\\
&=&V\tilde{\Ti}(\tilde{X} {\tilde{\rho}}^{-1} \tilde{Y})V^{\dagger}
=V\tilde{X} {\tilde{\rho}}^{-1} \tilde{Y}V^{\dagger}\\
&=&V\tilde{X}V^{\dagger}V {\tilde{\rho}}^{-1} V^{\dagger}V\tilde{Y}V^{\dagger}
=X {\rho}^{-1} Y.
\eeqan
Hence, it must be $\alg_\rho (W)\leq\fix(\Ti)$, as desired, and we still have $\supp(\alg_\rho(W))=\supp(\rho).$

On the other hand, if $X\in \textup{alg}_{\rho}(W)$,
then $\supp({\cal M}_{{\frac{1}{2}}}(X))\in\supp(W)$, and we have:
\beqan
 \Ti({\cal M}_{\rho^{\frac{1}{2}}}(X))& =&\Ti(VV^{\dagger}\rho^{\frac{1}{2}}X\rho^{-\frac{1}{2}}VV^{\dagger})
=VV^{\dagger}\Ti(V\tilde{\rho}^{\frac{1}{2}}\tilde X \tilde\rho^{-\frac{1}{2}}V^{\dagger})VV^\dagger\\
&=&V\tilde{\Ti}(\tilde{\rho}^{\frac{1}{2}}\tilde X {\tilde{\rho}}^{-\frac{1}{2}})V^{\dagger}
=V\tilde{\rho}^{\frac{1}{2}}\tilde X {\tilde{\rho}}^{-\frac{1}{2}}V^\dag\\
&=&{\rho}^{\frac{1}{2}}X {{\rho}}^{-\frac{1}{2}}.
\eeqan
Accordingly, ${\cal M}_{{\frac{1}{2}}}(W)\in\fix(\Ti)$ and $\supp({\cal M}_{{\frac{1}{2}}}(X))\subseteq\supp(\rho)$
as well, as desired. 
\qed
}

\subsection{From invariance to necessary conditions for stabilizability}

In order to apply the above lemma to our case of interest, namely, finding necessary conditions for FFQLS, 
the first step is to show that the reduced neighborhood states of $\rho$ may be used to generate the 
minimal $\rho$-distorted algebra containing the Schmidt span:

\begin{Prop}
\label{thm:redsupport}
Given a neighborhood $\neigh_j\in \neigh$, the support of the corresponding reduced state, $\rho_{\neigh_j}=\tr{\overline{\neigh}_j}{\rho}$, 
is equal to the support of the operator-Schmidt span $\Sigma_{\neigh_j}(\rho)$.
\end{Prop}

\vspace*{10pt}
\noindent
{\bf Proof:} 
{Since $\rho_{\neigh_j} \in\Sigma_{\neigh_j}  (\rho),$ $\text{supp}(\rho_{\neigh_j}) \leq \text{supp}(\Sigma_{\neigh_j}(\rho)).$ 
{\lv It remains to show the opposite inclusion, that is, by equivalently considering the complements}, 
that $\text{ker}(\rho_{\neigh_j})\leq \text{ker}(\Sigma_{\neigh_j}(\rho))$. Let $\ket{\psi}\in\text{ker}(\rho_{\neigh_j})$. Since $\rho_{\neigh_j}\geq 0$, we then have $\tr{}{\rho_{\neigh_j}\ketbra{\psi}}=\tr{}{\rho(\ketbra{\psi}\otimes\identity)}=0$. Let $\{E_i\}$ be a 
positive-operator valued measure (POVM) on $\hilbert_{\overline{\neigh}_j}$ which is informationally complete 
(that is, $\textup{span}\{E_i\} = \mathcal{B} (\hilbert_{\overline{\neigh}_j})$). 
The POVM elements sum to $\identity$, giving $\sum_i\tr{}{\rho(\ketbra{\psi}\otimes E_i)}=0$. Since each term is non-negative, $\tr{}{\rho(\ketbra{\psi}\otimes E_i)}=0$ for all $i$. Letting $\rho_i\equiv\tr{\overline{\neigh}_j}{\rho (\identity\otimes E_i)}$, we can write $0=\tr{}{\rho(\ketbra{\psi}\otimes E_i)}=\bra{\psi}\rho_i\ket{\psi}$.
Then, $\rho_i\geq 0$ implies $\rho_i\ket{\psi}=0$ for all $i$. Since the $E_i$ span the operator space $\mathcal{B}(\hilbert_{\overline{\neigh}_j})$, by using Eq. (\ref{eq:partialinnerprod}), we have that the corresponding $\rho_i$ span $\Sigma_{\neigh_j}(\rho)$. Hence, $\ket{\psi}\in\ker(\Sigma_{\neigh_j}(\rho))$.  \qed}

\vspace*{2mm}

\noindent 
The above Proposition, together with Lemma \ref{thm:localkernel} and Lemma 
\ref{thm:CPTPfix}, imply the following:

{\begin{cor}
\label{thm:crucialcoro}
If a state $\rho$ is in the kernel of a neighborhood generator 
$\lind_{j}={\lind}_{\neigh_j}\otimes\mathcal{I}_{\overline{\neigh}_j}$, then 
the minimal fixed-point set generated by the neighborhood Schmidt span obeys
\begin{equation}
{\cal F}_{\rho_{{\neigh}_j}}(\Sigma_{{\neigh}_j}(\rho))\otimes
\mathcal{B}(\hilbert_{\overline{\neigh}_j})\leq\ker(\lind_{_j}).
\end{equation}
\end{cor}

\noindent
 {\bf Proof:} 
Assume that $\rho\in\ker(\lind_{j})$. By Lemma \ref{thm:localkernel}, we have 
$\Sigma_{{\neigh}_j}(\rho)\otimes\mathcal{B}(\hilbert_{{\overline{\neigh}_j}})\leq\ker(\lind_{j}).$ 
By Proposition \ref{thm:redsupport}, we also know that the support of $\rho_{{\neigh}_j}$ is equal to that of 
$\Sigma_{{\neigh}_j}(\rho)$, and hence $\textup{supp}(\rho_{{\neigh}_j}\otimes\identity_{{\overline{\neigh}_j}})=\textup{supp}(\Sigma_{{\neigh}_j}(\rho)\otimes\mathcal{B}(\hilbert_{{\overline{\neigh}_j}}))$. With this and the fact that 
$$\rho_{{\neigh}_j}\otimes\identity_{{\overline{\neigh}_j}}\in\Sigma_{{\neigh}_j}(\rho)\otimes
\mathcal{B}(\hilbert_{{\overline{\neigh}_j}})\leq\ker(\lind_{j}),$$
Lemma \ref{thm:CPTPfix} implies that
$$ {\cal F}_{ \rho_{{\neigh}_j}\otimes\identity_{{\overline{\neigh}_j}}  } \Big( \Sigma_{{\neigh}_j}(\rho)\otimes\mathcal{B}(\hilbert_{{\overline{\neigh}_j}})  \Big)  \leq\ker(\lind_{j}),$$
or, equivalently,
${\cal F}_{ \rho_{{\neigh}_j}}( \Sigma_{{\neigh}_j}(\rho))\otimes\mathcal{B}(\hilbert_{{\overline{\neigh}_j}})
\leq\ker(\lind_j), $ as desired. \qed}

\vspace*{1mm}

Summing up the results obtained on invariance so far, and recalling that uniqueness of the equilibrium state is 
necessary for GAS, we have the following necessary condition:

\begin{thm}
\label{thm:neccond}
{\bf (Necessary condition for FFQLS)} 
A state $\rho$ is FFQLS relative to the neighborhood structure $\neigh$ only if {
\begin{equation}
\label{eq:nec1}
\textup{span}(\rho)=
\bigcap_j {\cal F}_{\rho_{{\neigh}_j}}(\Sigma_{{\neigh}_j}(\rho))\otimes
\mathcal{B}(\hilbert_{{\overline{\neigh}_j}}).
\end{equation}}
\end{thm}

\noindent
{\bf Proof:} {
Let $\rho$ be FFQLS relative to $\neigh$. Frustration-freeness of $\lind$ implies that 
$\ker(\lind)=\bigcap_{{\neigh}_j}\ker(\lind_{j})$. From QLS, we have $\ker(\lind)=\textup{span}(\rho)$. 
Thus, FFQLS implies 
$$\bigcap_j \ker(\lind_{j})=\textup{span}(\rho).$$
Corollary \ref{thm:crucialcoro} implies that for each neighborhood, ${\cal F}_{\rho_{{\neigh}_j}}(\Sigma_{{\neigh}_j}(\rho))\otimes\mathcal{B}(\hilbert_{{\overline{\neigh}_j}})\leq\ker(\lind_{j})$. Hence,
$$\bigcap_j {\cal F}_{\rho_{{\neigh}_j}}(\Sigma_{{\neigh}_j}(\rho))\otimes
\mathcal{B}(\hilbert_{{\overline{\neigh}_j}})\leq\bigcap_j \ker(\lind_{j}).$$
By construction, we also have
$$ \text{span}(\rho)\leq\bigcap_j {\cal F}_{\rho_{{\neigh}_j}}(\Sigma_{{\neigh}_j}(\rho))\otimes\mathcal{B}(\hilbert_{{\overline{\neigh}_j}}).$$
Stringing together the three relationships above, we arrive at the desired result, Eq. (\ref{eq:nec1}).}
\qed

\section{Frustration-Free Stabilizable States: Sufficient Conditions}
\label{sec:sufficiency}

In this section, we move from necessary conditions for FFQL stabilization to sufficient ones, by 
providing in the process a {\em constructive} procedure to design stabilizing semigroup generators. 
A key step will be to establish a property that {\em arbitrary} (convex) sums of Liouvillians enjoy, namely the fact that, 
{as long as the algebras associated with individual components of the generator are contained in the algebra associated to the full generator, the existence of a {\em common full-rank fixed point} suffices to prove frustration-freeness.
Drawing on this result, we will prove that the necessary condition of the previous section is also sufficient in the generic case where the target state is full-rank, and then separately address general target states. }

\subsection{A key result on frustration-free Markovian evolutions}
\label{sec:fflindblads}

Consider a QDS of the form $\Li=\sum_k\Li_k$, where individual terms need {\em not}, at this stage, 
correspond to neighborhood generators. The following general result holds: 

\begin{thm}
\label{thm:frustration-new}
{\bf (Common fixed points of sums of Liouvillians) }
Let $\lind=\sum_{k}\lind_{k}$ be a sum of QDS generators, and assume that  
the following conditions hold: \\
\noindent (i) $\alg\{\lind_k\}\leq\alg\{\lind\}$ for each $k$; \\
\noindent (ii) there exists a positive definite $\rho\in\ker(\Li)$ such that $\rho\in\ker(\lind_k)$ for all $k$.\\
\noindent 
Then $\rho'$ is invariant under $\lind$ only if it is invariant under all $\lind_k$, that is:
\[\rho'\in\ker(\lind) \implies \rho'\in\ker(\lind_k) \quad \forall\,k .\] 
\end{thm}

\noindent
{\bf Proof:} 
By linearity of $\lind$, we clearly have that $\ker(\lind)\geq \bigcap_k\ker(\lind_k)$. We show that under 
the hypotheses, $\ker(\lind)\leq\bigcap_k\ker(\lind_k)$, therefore effectively implying 
$\ker(\lind)=\bigcap_k\ker(\lind_k)$.
By (ii), $\rho$ is a full-rank state in $\ker(\lind)$ and $\rho\in \ker(\lind_k)$ for all $k$.
Theorem \ref{thm:dualkernel} implies that
\beqan
 \ker(\lind)= \rho^\frac{1}{2}\ker(\lind^\dag)\rho^\frac{1}{2} \quad 
\textup{and} \quad
\ker(\lind_k)=\rho^\frac{1}{2}\ker(\lind_k^\dag)\rho^\frac{1}{2}, \:\forall k.
\eeqan
Then, by Lemma \ref{thm:commutantkernel}, we also have that
\beqan
\rho^\frac{1}{2} \ker(\lind^\dag) \rho^\frac{1}{2} &=& \rho^\frac{1}{2} \alg\{\lind\}'  \rho^\frac{1}{2}
\quad \textup{and}\quad 
\rho^\frac{1}{2} \ker(\lind_k^\dag) \rho^\frac{1}{2} = \rho^\frac{1}{2} \alg\{\lind_k\}' \rho^\frac{1}{2} .
\eeqan
In view of condition (i), the relevant commutants satisfy 
\[ \alg \{\lind\}'\leq \alg\{\lind_k\}' , \quad \forall k. \]
\noindent 
The above inequality may then be used to bridge the previous equalities, yielding\vspace*{1mm}:
\begin{center}
\begin{tabular}{ccccc}
$\ker(\lind)$ & $=$ & $\rho^\frac{1}{2}\ker(\lind^\dag)\rho^\frac{1}{2}$ & $=$ & $\rho^\frac{1}{2}\alg 
\{\lind\}' \rho^\frac{1}{2}$ \\
&  &  &  & \vspace*{-2mm}\\
&  &  &  & $\text{\rotatebox[origin=c]{-90}{$\leq$}}$ \vspace*{-2mm}\\
&  &  &  &  \\
$\ker(\lind_k)$ & $=$ & $\rho^\frac{1}{2}\ker(\lind_k^\dag) \rho^\frac{1}{2} $ & $=$ & $\rho^\frac{1}{2} 
\alg\{\lind_k\}' \rho^\frac{1}{2}$,
\end{tabular}\vspace*{1mm}
\end{center}
for all $k$. From this we obtain $\ker(\lind)\leq \bigcap_k\ker(\lind_k)$, which completes the proof.
\qed

\vspace*{1mm}
\noindent 
{\bf Remark 3.}
We note that condition (i) above, namely $\alg \{\lind_k\}\leq\alg\{\lind\},$ is only ever
{\em not} satisfied due to the presence of Hamiltonian contributions in $\lind_k$. 
In fact, if $\lind_k\equiv \lind_k(\{L_{j,k}\})$ for each $k$ in a given representation, then 
$\lind=\sum_k\lind_k$ also has a purely dissipative representation $\lind(\bigcup_k\{L_{j,k}\}),$ and thus
$\alg\{\lind_k\}\leq\alg\{\lind\}.$ 
On the other hand, suppose that $\lind_k=\lind_k(H_k,\{L_{j,k}\}),$ with $H_k\ne0$ in some representation. 
This implies that $\lind =\lind(H,\bigcup_k\{L_{j,k}\}),$ with $H=\sum _k H_k$. In this case, since $\alg(H)$ need not contain 
$\alg(H_k),$ condition (i) does not hold in general.  As a trivial example, consider two generators associated to 
$H_1=M$ and $H_2=-M,$ with $M\neq {\mathbb I}.$ Clearly, $\{{\mathbb O}\}=\alg(H)$ does not contain $\alg(M).$ 
Likewise, if $\Hi =({\mathbb C}^2)^{\otimes 3}$ and $Z$ is a single-qubit Pauli operator, consider QL Hamiltonians 
$H_1 = ZZ {\mathbb I}$ and $H_2 = {\mathbb I}ZZ$. Then $\alg(H_1) \nleq \alg(H_1+H_2)$. 
Intuitively, this stems from the fact that since noise operators enter ``quadratically'' (bilinearly) in the QDS,
they cannot cancel each other's action -- unlike Hamiltonians, which by linearity 
may ``interfere'' with one another.

{Interestingly, the reasoning leading to Theorem \ref{thm:frustration-new} also applies to CPTP maps, with the 
simplification that, since no Hamiltonian is present, condition (i) is always satisfied. Formally:}

\begin{cor}
\label{cor:frustrationkraus}
{\bf (Common fixed points of sums of CPTP maps)}
Let $\Ti=\sum_{k} p_k\Ti_{k}$ be a sum of CPTP maps, with $p_k>0$ and $\sum_k p_k=1$. 
If there exists a positive definite $\rho \in \fix(\Ti)$ such that $\rho\in\fix(\Ti_k)$ for all $k$, then 
$\rho'$ is invariant under $\Ti$ only if it is invariant under all $\Ti_k$, that is:
 \[\rho' \in\fix(\Ti) \implies \rho' \in\fix(\Ti_k). \] 
\end{cor}
\noindent 
The proof is a straightforward adaptation of the one above, and it can actually be extended to a positive {\em semi-definite} fixed state $\rho$, provided that some extra hypotheses on the support of $\rho$ are satisfied -- 
see Appendix B for a precise statement and its proof. 

Another direct corollary of Theorem \ref{thm:frustration-new}, which now specializes to {\em locality-constrained dynamics}, 
provides us with a useful tool to ensure that a QL generator be FF: the generator itself and {\em all} of its QL 
components must share a full-rank fixed state. 

\begin{cor}
\label{thm:frustration}
{\bf (Frustration-freeness from full-rank fixed point)}
Let $\lind=\sum_{j}\lind_{j}$ be a QL generator, and assume that  
the following conditions hold: \\
\noindent (i) $\alg\{\lind_j\} \leq \alg\{\lind\}$ for each $j$; \\
\noindent (ii) there exists a positive-definite $\rho\in\ker(\Li)$ such that $\rho\in\ker(\lind_j)$ for all $j$.\\
\noindent 
Then the QL generator $\Li$ is FF. 
\end{cor}

\subsection{Sufficient conditions for full-rank target states}
\label{sec:suffcond-generic}

\label{subsec:fullrank}
\begin{thm}
\label{thm:suffcond}
{\bf (Sufficient condition for full-rank FFQLS)} 
A full-rank state $\rho$ is FFQLS relative to the neighborhood structure $\neigh$ if {
\begin{equation}
\textup{span}(\rho)=
\bigcap_j {\cal F}_{\rho_{{\neigh_j}}}(\Sigma_{{\neigh_j}}(\rho))
\otimes\mathcal{B}(\hilbert_{\overline{\neigh}_j}).
\label{hypo}
\end{equation}}
\end{thm}

\noindent
{\bf Proof:} 
To show that this condition suffices for FFQLS, we must show that there exists some QL FF Liouvillian $\lind$ 
for which $\textup{span}(\rho)=\ker(\lind)$. Our strategy is to first construct a QL generator for which $\rho$ is 
the unique state in the intersection of the QL-components' kernels. Then, we use Thm. \ref{thm:frustration} to 
show that this generator is FF, yielding the desired equality.

Fix an arbitrary neighborhood $\neigh_j \in \neigh$, with associated bipartition 
$\Hi = \Hi_{\neigh_j} \otimes \Hi_{\overline{\neigh}_j}$. 
We shall construct a neighborhood CPTP map ${\cal E}_j \equiv {\cal E}_{\neigh_j} \otimes {\cal I}_{\overline{\neigh}_j}$, 
where ${\cal E}_{\neigh_j}$ {{\em projects onto the minimal fixed-point set containing the neighborhood-Schmidt 
span} (that is, such projection maps are duals to a conditional expectation).
{Since, by construction, ${\cal F}_{\rho_{{\neigh_j}}}(\Sigma_{{\neigh_j}}(\rho))$ is a 
modular-invariant distorted subalgebra of ${\cal B}(\Hi_{\neigh_j})$,
Theorem \ref{thm:modinvariance} ensures that there exists a CPTP map $\Ei_{{\neigh_j}}$ such that
\[\fix(\Ei_{\neigh_j})={\cal F}_{\rho_{{\neigh_j}}}(\Sigma_{{\neigh_j}}(\rho)).\]  
In particular, we take
$\Ei_{{\neigh_j}}^2=\Ei_{{\neigh_j}},$ so that it projects onto its fixed points.}
Explicitly, its structure follows from the decomposition in Eq. \eqref{CPTPfixedpoints}:
$${\cal F}_{\rho_{{\neigh_j}}}(\Sigma_{{\neigh_j}}(\rho)) =
\bigoplus_\ell \mathcal{B}(\hilbert_{\ell,j}^{(A)}) \otimes \tau_{\ell,j}^{(B)} , $$
with a corresponding Hilbert space decomposition 
$\Hi_{\neigh_j} \equiv \bigoplus_\ell  {\Hi}_{\ell,j} = \bigoplus_\ell \Hi_{\ell,j}^{(A)}\otimes\Hi_{\ell,j}^{(B)},$ and 
$\tau_{\ell, j}^{(B)}$ a full-rank state on $\Hi_{\ell,j}^{(B)}$.
{Introducing partial isometries $\Pi_{\ell,j}: \Hi_{\ell,j} \rightarrow \Hi_{\neigh_j}$,
the sought-after maps ${\cal E}_{\neigh_j}$ can be constructed as:
\begin{equation}
\label{eq:dualcondex}
\mathcal{E}_{{\neigh_j}}(\rho) \equiv \bigoplus_\ell \Tr_{\Hi_{\ell,j}^{(B)}} (\Pi_{\ell,j}^{\dagger} \rho\Pi_{\ell,j}) 
\otimes \tau_{\ell, j}^{(B)}. 
\end{equation}
It is straightforward to verify that $\mathcal{E}_{{\neigh_j}}(\rho)$ is CPTP.}}
Recalling Eq. (\ref{lindbladize}), we may then define a neighborhood QDS generator by taking 
$\kappa = {\cal E}_{\neigh_j}^\dag ({\mathbb I})/2 = {\mathbb I}/2$ and letting 
\begin{equation}
\label{eq:neighlind}
\lind_{{\neigh_j}} \equiv \mathcal{E}_{{\neigh_j}} - {\cal I}_{{\neigh}_j} , \quad \forall j . 
\end{equation}

Let now $\lind \equiv \sum_j \lind_j = \sum_j \lind_{{\neigh_j}} \otimes {\cal I}_{\overline{\neigh}_j}$ 
define the QL generator of the overall dynamics. We constructed each $\lind_j$ in such a way that 
$$\ker(\lind_j) = {\cal F}_{\rho_{{\neigh_j}}}(\Sigma_{{\neigh_j}}(\rho))\otimes\mathcal{B}(\hilbert_{\overline{\neigh}_j}), 
\quad \forall j. $$
Hence, by invoking the hypothesis (Eq. (\ref{hypo})), it follows that $\rho$ is the unique state obeying 
\begin{equation}
\textup{span}(\rho)=\bigcap_j \ker(\lind_{{j}}) \leq \ker(\lind).
\label{k0}
\end{equation}
{A priori, it is still possible that 
$\textup{span}(\rho) < \ker(\lind)$. However, since we have chosen $\kappa=\kappa^\dag$, 
the neighborhood generators $\lind_j$ defined in Eq. (\ref{eq:neighlind}) do {\em not} have any 
Hamiltonian contribution; recalling 
Remark 3, it follows that the algebra of the global generator contains the algebra of each 
neighborhood generator,}
$$\alg\{\lind\} \geq \alg\{\lind_j \}, \quad \forall j. $$
Thus, by Corollary \ref{thm:frustration}, the generator $\lind$ is FF. 
From $\lind$ being FF, it follows in turn that
$$\ker(\lind) \leq \bigcap_j \ker(\lind_{j}),$$
which, together with Eq. (\ref{k0}), implies $\textup{span}(\rho)=\ker(\lind)$, as desired.
\qed

\subsection{Sufficient conditions for general target states}
\label{subsec:suffcond}

If the target state $\rho$ is not full-rank, the necessary condition of Theorem \ref{thm:neccond} may still be 
shown to be sufficient for FFQLS if an additional condition (referred to as the ``support condition'' henceforth) 
is also obeyed:

\begin{thm}
\label{thm:suffcondnorank}
{\bf (Sufficient condition for general FFQLS)} 
An arbitrary state $\rho$ is FFQLS relative to the neighborhood structure $\neigh$ if {
\begin{equation}
\textup{span}(\rho)=
\bigcap_j {\cal F}_{\rho_{\neigh_j}}(\Sigma_{\neigh_j}(\rho))\otimes\mathcal{B}(\hilbert_{\overline{\neigh}_j})
\end{equation}}
and
\begin{equation}
\label{eq:suppcond}
\textup{supp}(\rho)=\bigcap_j \textup{supp}(\rho_{\neigh_j}\otimes\identity_{\neigh_j}).
\end{equation}
\end{thm}

\vspace*{10pt}
\noindent
{\bf Proof:} 
Our strategy is to use the support condition of Eq. (\ref{eq:suppcond}) to reduce the non-full-rank case to the full-rank one. 
As in the proof of the previous theorem, fix an arbitrary neighborhood $\neigh_j$, and consider the maps ${\cal E}_{\neigh_j}$, defined in Eq. (\ref{eq:dualcondex}).  Let $P_{\neigh_j} \in {\mathcal B}(\Hi_{\neigh_j})$ denote the Hermitian projector onto {$\text{supp}({\cal F}_{\rho_{\neigh_j} } (\Sigma_{\neigh_j}) (\rho))$}, and $P^{\perp}_{\neigh_j}= {\mathbb I}_{\neigh_j} -P_{\neigh_j}$ the associated orthogonal projector. 
In this case, we compose each ${\cal E}_{\neigh_j}$ with the corresponding map
\begin{equation}
\mathcal{E}_{\neigh_j}^0(\cdot)  \equiv P_{\neigh_j} (\cdot ) P_{\neigh_j} + \frac{ P_{\neigh_j} }{ \Tr ({P_{\neigh_j}}) }  \,
\Tr \,( P^{\perp}_{\neigh_j}\cdot ),   
\label{newmaps}
\end{equation}
where $\mathcal{E}_{\neigh_j}^0$ is, like $\mathcal{E}_{\neigh_j}$, both CP and TP:
$$ \Tr \,( { \mathcal{E}_{\neigh_j}^0 (M)} ) = \Tr \Big( { M(P_{\neigh_j} + P^{\perp}_{\neigh_j})} \Big) = \Tr ({M}), 
\quad \forall M \in  {\mathcal B}(\Hi_{\neigh_j}).$$ 
\noindent
With this, consider new CPTP maps given by $ \mathcal{E}_{\neigh_j}\circ\mathcal{E}_{\neigh_j}^0, $
whereby it follows that new neighborhood generators may be constructed as 
\begin{equation}
\lind_{\neigh_j} \equiv \mathcal{E}_{\neigh_j}\circ\mathcal{E}_{\neigh_j}^0-\mathcal{I}_{\neigh_j},  \quad 
\lind_j = \lind_{\neigh_j} \otimes {\cal I}_{\overline{\neigh}_j},
\label{lindj}
\end{equation}
with the global evolution being driven, as before, by the QL generator ${\cal L}=\sum_{j}{\cal L}_{j}$.
 
Define now $\Pi$ to be the projector onto $\textup{supp}(\rho)$, and consider the 
positive-semidefinite function
$V(\tau)=1-\tr{}{\Pi \, \tau},$ $\tau \in {\mathcal B}(\Hi).$
The derivative of $V$ along the trajectories of the generator we just constructed is
\[\dot V(\tau)=-\sum_j\tr{}{\Pi \, {\cal L}_j(\tau)}.\]
By LaSalle-Krasowskii theorem \cite{khalil-nonlinear}, the trajectories will converge to the largest invariant set 
contained in the set of $\tau$ such that the above Lyapunov function $\dot V(\tau)=0$. We next show that this set 
must have support {\em only} on $\textup{supp}(\rho)=\bigcap_{\neigh_j}\textup{supp}(\rho_{\neigh_j}\otimes 
{\identity}_{\overline{\neigh}_j})$. 
{Since $V$ is defined on global input operators, we first re-express each neighborhood 
generator $\lind_j$ in Eq. (\ref{lindj}) as }
\[ \lind_j = {\cal E}_j \circ {\cal E}_j^0 - {\mathcal I}, \quad \quad 
{\cal E}_j \equiv {\cal E}_{\neigh_j} \otimes {\cal I}_{\overline{\neigh}_j}, \quad
{\cal E}_j^0 \equiv {\cal E}_{\neigh_j}^0 \otimes {\cal I}_{\overline{\neigh}_j},\]
\noindent 
where we have used the property ${\cal E}_j \circ {\cal E}_j^0=  ({\cal E}_{\neigh_j} \circ {\cal E}^0_{\neigh_j})\otimes 
{\cal I}_{\overline{\neigh}_j}$. Additionally, let $P_j\equiv P_{\neigh_j} \otimes {\mathbb I}_{\overline{\neigh}_j}$ 
denote the projector onto {$\text{supp}( {\cal F}_{\rho_{\neigh_j} } (\Sigma_{\neigh_j} (\rho)) \otimes {\mathbb I}_{\overline{\neigh}_j})$.}
Assume now that $\textup{supp}(\tau)\nsubseteq \textup{supp}(\rho_{\neigh_k}\otimes\identity_{\overline{\neigh}_k})$ for some $\neigh_k \in \neigh$, that is,  $\Tr{}{ (\tau P_k^\perp )} >0$. 
By using the explicit form of the maps ${\cal E}_{\neigh_j}^0$ given in Eq. (\ref{newmaps}), we then have:
\begin{eqnarray}
\dot V(\tau)&\leq & - \tr{}{\Pi \, {\cal L}_{k}(\tau)}\nonumber \\
&=&-\tr{}{\Pi \,(\mathcal{E}_k \circ\mathcal{E}_k^0 ) \tau  - \Pi \, \tau} \nonumber\\
&=&-\tr{}{\Pi \mathcal{E}_k (P_k \tau P_k ) } - \tr{}{\tau P_k^\perp} 
\frac{ \tr{}{\Pi \mathcal{E}_k (P_k)} }{  \tr{}{P_k }  }   + \tr{}{\Pi\tau} .
\label{eq:third}
\end{eqnarray}
Since the target state $\rho$ is invariant under ${\cal E}_k$, its support is also invariant. This implies that
\[\tr{}{\Pi \mathcal{E}_k (P_k \tau P_k )} \geq \tr{}{\Pi\mathcal{E}_k (\Pi \tau \Pi)}=\tr{}{\Pi\tau}.\]
\noindent
Hence, the sum of the first and the third term in Eq. \eqref{eq:third} is less than or equal to zero.
The second term, on the other hand, is {\em strictly negative}. This is because: (i) we assumed that 
$\Tr ( {\tau P_k^\perp  } ) >0$;  (ii) with $\Pi\leq P_k,$ and ${\cal E}_k (P_k)$ 
having the same support of $P_k $ by construction, it also follows that
{$\tr{}{\Pi {\cal E}_k (P_k )}>0.$}
We thus showed that no state $\tau$ with support outside of the support of $\rho$ can be in the attractive set for the dynamics. 
Hence, the dynamics asymptotically converges onto the support of $\rho,$ which is invariant for all the $\lind_j$. 
By restricting to this set, the maps ${\cal E}_j^0$ have no effect and the same argument of Theorem \ref{thm:suffcond} 
shows that the only invariant set in such a subspace is $\textup{span}(\rho)$, as desired.
\qed

\vspace*{1mm}

\noindent
{\bf Remark 4.} 
We note that the support condition in Eq. \eqref{eq:suppcond} is indeed a natural candidate 
for a sufficient FFQLS condition, since if $\rho$ is pure, it reduces to the necessary and 
sufficient condition of Theorem \ref{mainthm}.
However, it is provably {\em not} necessary in general, see Sec. \ref{sec:suppcounter}.
{\em We conjecture that the necessary condition of Theorem \ref{thm:neccond} is in fact sufficient for 
general (non-full-rank) states}. However, we currently lack a complete proof.

\section{Illustrative Applications}
\label{sec:appl}

In this section, we illustrate the general framework presented thus far through a number of examples, 
with a twofold goal in mind: to both demonstrate the applicability and usefulness of the mathematical tools we 
have developed, and to gain insight into the problem of {\em mixed-state} QL stabilization, along with 
appreciating important differences from the pure-state setting.
For simplicity, we shall focus in what follows on a multi-partite system consisting of $n$ co-dimensional qudit
subsystems, namely, $\Hi= \otimes_{a=1}^n \Hi_a = ({\mathbb C}^d)^{\otimes n}$, with $D= \text{dim}(\Hi)=d^n$. 
In the especially important 
case corresponding to qubit (or spin-$1/2$) subsystems, $d=2$, we shall follow standard notation and denote by 
$\{ \ket{0}, \ket{1}\}$ an orthonormal (computational) basis in ${\mathbb C}^2$ and by 
$\{ \sigma_\alpha |\alpha =0,1,2,3\} \equiv \{ {\mathbb I}, X, Y, Z\}$ the set of single-qubit Pauli matrices, under the 
natural extension to multi-qubit operators, e.g., $\sigma_x^{(a)} \equiv X_a ={\mathbb I} \otimes \ldots 
{\mathbb I}\otimes X \otimes \ldots {\mathbb I}$, with non-trivial action occurring only on the $a$th factor.

\subsection{Some notable failures of quasi-local stabilizability}
\label{failures}

Before exhibiting explicit classes of states which are provably FFQLS, it may be useful to appreciate some distinctive features that the mixed nature of the target state entails and, with that, the failure of some intuitively natural mechanisms to generate candidate FFQLS states. Recall that an arbitrary pure product (fully factorized) state is always DQLS (or equivalently, as shown, FFQLS) \cite{ticozzi-ql} thus, in other words, failure of a pure target state to be FFQLS always implies some entanglement in the state. In contrast to that, 
{\em entanglement is not necessary for failures of FFQLS if the target is mixed.}
Consider $n$ qubits arranged on a line, with neighborhood specified by nearest-neighbor (NN) pairs, 
$\neigh_j \equiv \{ j, j+1\}$, $j=1, \ldots, n-1$, and the manifestly separable target state 
\begin{equation} \rho \equiv \rho_\text{sep} = \frac{1}{2} ( \ket{0}\bra{0}^{\otimes n} +\ket{1}\bra{1}^{\otimes n}). 
\label{rhosep}
\end{equation}
\noindent 
Because $\rho$ is already in Schmidt decomposition form for all $n$, it easily follows that each Schmidt span
has the form
$$ \Sigma_{\neigh_j}(\rho) = \text{span} 
\{ \ket{00}\bra{00}_{j,j+1} ,
\ket{11}\bra{11}_{j,j+1} \}, $$
and is {already a $\rho_{\neigh_j}$-distorted algebra invariant for ${\cal M}_{{\frac{1}{2}}}$.} 
Taking the intersection over all neighborhoods then 
leaves the two-dimensional space,
$$ \bigcap_j  \Sigma_{\neigh_j}(\rho) \otimes {\mathcal B}(\Hi_{\overline{\neigh}_j} ) = 
\span \{ \ket{0}\bra{0}^{\otimes n}, \ket{1}\bra{1}^{\otimes n} \} > \text{span}(\rho), $$
which violates the necessary condition for FFQLS of Theorem \ref{thm:neccond}.

Likewise, mixing a pure FFQLS entangled state $\ket{\psi}$ with a trivially FFQLS target such as the fully 
mixed state results in a ``pseudo-pure'' target state of the form 
\begin{equation}
\rho\equiv \rho_\text{pp} = (1-\epsilon) \ket{\psi}\bra{\psi} + \epsilon \, {\mathbb I}/2^n, 
\label{rhopp}
\end{equation}
which is {\em not FFQLS} in general: an explicit example may be constructed by taking $\ket{\psi}$ to be 
the two-excitation Dicke state on $n=4$ qubits, 
\begin{equation}
D_{4,3}\equiv\ket{(0011)} = \frac{1}{\sqrt{6}} ( \ket{0011} +\ket{0101} + \ket{0110} + \ket{1001}+\ket{1010} +\ket{1100}),
\label{D4}
\end{equation}
which was proved to be DQLS relative to the three-body neighborhoods $\neigh_1 = \{ 1,2,3 \}$, $\neigh_2=\{ 2,3,4\}$
in \cite{ticozzi-ql} (see also Sec. \ref{Dicke} and Appendix C for explicit calculations).

\subsection{Quasi-local stabilization of graph product states}
\label{sub:graph}

Multi-qubit pure graph states are an important resource across QIP, with applications ranging from 
measurement-based quantum computation \cite{Briegel2001} to stabilizer quantum error-correcting codes \cite{Schlingemann}. 
More recently, thermal graph states \cite{Kay2006,Temme2014} have been shown to both provide faithful approximations of pure graph states for sufficiently low temperatures and to support non-trivial multipartite bound entanglement over a temperature range \cite{Cavalcanti2010}. In this section, we demonstrate that a broader class of mixed graph states on qudits, which we refer 
to as {\em graph product states}, are FFQLS.

For the special case of qubits, both pure \cite{Kraus2008,ticozzi-ql} and thermal \cite{Temme2014} graph states are known to be stabilizable with QL FF semigroup dynamics with respect to a natural  locality notion induced by the graph. We recover and extend these results to $d>2$ and a broader class of non-thermal graph states, without making reference, in the mixed-state case, to properties of the Davies QDS generator which is typically employed under weak-coupling-limit assumptions 
\cite{alicki-lendi,Brandao2014,Temme2014}.
The key property of graph product states is that they can be transformed to a product form relative to a 
``logical subsystem factorization,'' following a change of basis which is effected by a sequence of 
commuting neighborhood unitary transformations (a QL quantum circuit). 
Commutativity of the unitaries effectively reduces the problem of FFQLS to one of local stabilization of product states. 
While for graph states this observation allows to directly obtain QL FF stabilizing dynamics, they nevertheless serve as a 
first relevant example of FFQLS, in preparation for cases where the tools we propose become indispensable. 

We formally define general {\em qudit graph states} following \cite{Zeng2015}. Let $G=(V,E)$ be a graph, 
where vertices $j\in V$ are associated to qudits and edges $(j,k)\in E$ label their allowed pairwise interactions. 
A natural neighborhood structure is derived from $G$, by letting the $j$th neighborhood $\neigh_j$ comprise vertex $j$ along with the subset of vertices adjacent to it. 
Rather than associating the graph to a particular state, the graph is used to construct a set of commuting unitary edge-wise operators, say $\{U_{(j,k)}\}$. 
The product of all such unitaries, ${U}_G \equiv\prod_{(j,k)\in E} U_{(j,k)}$, constitutes the quantum circuit 
which is used to map an input product state into, in general, an entangled 
state\footnote{In the cluster model of quantum computation, 
qubit graph states are constructed by applying this global unitary (as a sequence of 
commuting neighborhood-wise actions) on an initial product state $\ket{+}^{\otimes n}$.}.
Each $U_{(j,k)}$ is defined as the generalized controlled-$Z$ transformation \cite{Zeng2015} associated to a symmetric qudit Hadamard matrix $H$.
Hadamard matrices $H$ are defined by the conditions $H^{\dagger}H=d\identity$, $H=H^T$, and $|h_{ij}|=1$, where $h_{ij}\equiv[H]_{ij}$. 
To each Hadamard, there exists a corresponding generalized controlled-$Z$ gate acting on two-qudits, defined by $ C^H\ket{ij}=h_{ij}\ket{ij}$. 
With these in place, $U_{(j,k)} \equiv C^H_{(j,k)}\otimes\identity_{\overline{(j,k)}}$, and 
the unitary transformation which transforms local operators to neighborhood operators on
$\neigh_j$ is defined by
\begin{equation*}
U_j\equiv \prod_{k\in \neigh_j\backslash j} C^H_{(j,k)},
\end{equation*}
where each operator is defined on the global Hilbert space $\Hi$, and acting non-trivially only on the 
subsystems by which it is indexed\footnote{ 
Since $H$ is not uniquely defined, the above 
$U_{(j,k)}$ depend on the choice of $H$. For readability, our notation does not make this explicit. For standard qubit graph states, $H$ is the discrete Fourier transform on ${\mathbb C}^2$.}.
Standard pure qudit graph states may be defined as \cite{Zeng2015} 
$\ket{\psi_{G}}\equiv {{U}_G}\, \ket{+}^{\otimes n}. $
Similarly, we define \emph{graph product states} as
\begin{equation}
\rho_{G}\equiv {{U}_G} \,\rho_\text{prod} \,{{U}_G}^\dagger 
={{U}_G} \Big(\bigotimes_{j=1}^n \rho_j \Big) {{U}_G}^{\dagger},
\label{rho_G}
\end{equation}
where each $\rho_j$ is an arbitrary qudit mixed state. We note that graph product states are distinct from 
(though overlapping with) so-called \emph{graph diagonal states} \cite{Briegel2005}, defined as those 
states obtained by applying the circuit ${{U}_G}$ to any state $\rho_{\textup{diag}}$ diagonal in the 
eigenbasis of the Hadamard matrix.
 
To construct QL Lindblad operators which stabilize a graph product state $\rho_G$, we may simply construct the local 
Lindblad operators which prepare each factor $\rho_j$ of Eq. (\ref{rho_G}) in the un-rotated basis, and then transform these Lindblad operators with ${{U}_G}$. Let each factor $\rho_j$ be diagonalized by $\rho_j = V_j (\sum_i \gamma_j^i \, \ketbra{i}) V_j^{\dagger}$, where $\gamma_j^i \geq0$ are the ordered (with $i$) eigenvalues of the qudit density operator $\rho_j$ and $V_j$ the diagonalizing unitary transformation. 
Stabilizing Lindblad operators may then be constructed as follows:
\begin{eqnarray*}
L^j_{i,i+1} = \sqrt{\gamma^j_i} \,U_j  V_j  \ket{i}\bra{i+1}V_j^{\dagger} U_j^{\dagger}, \quad
L^j_{i+1,i} =  \sqrt{\gamma^j_{i+1}} \, U_j V_j \ket{i+1}\bra{i}V_j^{\dagger} U_j^{\dagger}, 
\end{eqnarray*}
where $i,j =1, \ldots,n$ and each Lindblad operator is defined on the whole $\Hi$, but by construction acts non-trivially only on the neighborhood $\neigh_j$. That the resulting global dynamics $\lind = \sum_j \lind_j$ are FF 
follows from the commutativity of the neighborhood-Liouvillians $\lind_j$.

It is interesting to note that, since any {\em pure} state $\rho_\text{diag}$ in Eq. (\ref{rho_G}) which is diagonal in the computational basis is necessarily a product, {\em arbitrary pure qudit graph states are FFQLS}.  In general, however, 
since $\rho_\text{diag}$ may be separable but not necessarily of product form, {\em mixed graph diagonal states need not be FFQLS}  (in line with similar conclusions for trivially separable states, as discussed in Sec. \ref{failures}). 

\vspace*{1mm}

\noindent 
{\bf Remark 4: Graph Hamiltonians.}
Pure graph states may be equivalently defined as a special class of stabilizer states, by assigning to each vertex in $G$ a stabilizer generator, taken from the generalized Pauli group ${\cal G}_n$ for $n$ qudits \cite{Schlingemann}. For qubits, for example, a graph state $\ket{\psi_G}$ may be seen to the the unique ground state of a QL graph Hamiltonian $H_G$ 
that is a sum of generators of ${\cal G}_n$ of the form: 
\begin{equation}
H_G \equiv \sum_{j=1}^n H_{G,j} = - \sum_{j\in V} X_j \hspace*{-1mm}\bigotimes_{k \in {\neigh_j} \backslash j} 
\hspace*{-1mm} Z_k = -
{{U}_G}^\dagger \Big( \sum_j X_j \Big)\, {{U}_G}.
\label{Hgraph}
\end{equation}
By construction, $H_G$ is a sum of \emph{commuting} terms, and may be easily seen to be FF (namely, such that 
$\ket{\psi_G}$ is also the ground state of each $H_{G,j}$ separately).
Further to that, the last equality in Eq. (\ref{Hgraph}) makes it clear how the graph Hamiltonian is mapped to a (strictly) local one in the 
``logical basis'', following application of the circuit $U_G$.
A feature that becomes evident from expressing graph-Hamiltonians in this form, and that is {\em not} shared 
by more general QL commuting Hamiltonians, is the {\em large degeneracy} of their eigenspaces -- precisely $1/d$
of the global space dimension\footnote{For $d=2$, 
this feature is key in enabling graph-state preparation in finite time with discrete-time dynamics designed 
via splitting-subspace approaches \cite{Baggio2012}.}.
Thermal graph states \cite{Kay2006,Temme2014}, relative to Hamiltonians as in Eq. (\ref{Hgraph}), 
are a special case of graph product states, corresponding to each qudit being in a canonical Gibbs state, 
namely, $\rho_j \propto \exp(-\beta_j H_{G,j})$ in Eq. (\ref{rho_G}) 
(or, $\rho_G \propto U_G^\dag ( \otimes_j e^{\beta_j X_j} ) {U}_G$), where 
$\beta_j$ denotes the inverse equilibrium temperature of the $j$th qubit. 
Thermal qubit graph states can thus provide a scalable class of mixed multiparty-entangled states. 

The construction leading to graph product states may be generalized to arbitrary situations where a quantum circuit 
arising from commuting unitary neighborhood operators may be identified, not necessarily stemming from a graph. 
That is, say that ${U}\equiv \prod_{(j,k)\in E} U_{(j,k)}$, with each $U_{(j,k)}$, as above, being an edge-wise operator, with $[U_{(j,k)},U_{(j',k')}]=0$ for all edges $(j,k),(j',k')$.
Then we may define QL-transformed 
product states as resulting from the action of ${\cal U}$ on any product input state:
$\rho \equiv {U} \,\rho_\text{prod} \,{U}^\dagger.$
Note that $\rho_\text{prod}$ is, clearly, FFQLS in the strongest sense, relative to strictly local 
(single-site) neighborhoods, 
whereas $\rho$ is FFQLS relative to the structure $\{ \neigh_j = \bigcup_{j\cap (i,k) \neq \emptyset} (i,k)\}$, which is  
imposed by the circuit. More generally, a QL commuting circuit may be used to 
extend neighborhoods of some input ${\cal N}_{{\rm in}}$ into larger neighborhoods of some
output ${\cal N}_{{\rm out}}$, associated to weaker QL constraints. 
FFQLS states maintain their property under this type of  transformation, in the following sense:

\begin{Prop}
\label{QLComp}
{\bf (Circuit-transformed FFQLS) }
Let $\rho_{\rm in}$ be FFQLS relative to $\neigh_{{\rm in}} \equiv \{\neigh_{{\rm in},i} \}$, and let 
$\,{U}= \prod_j U_j$, with  $U_j \equiv U_{\neigh_j}\otimes\identity_{\overline{\neigh}_j}$ and 
$[U_j, U_{j'}]=0$ for all $j,j'.$  Then the output 
state $\rho_{\rm out}=\mathcal{U}(\rho)\equiv  {U}\rho_{\rm in}{U}^\dagger$ is FFQLS relative to  
$\neigh_{\rm out} \equiv \{\neigh_{{\rm out},k}  \}$, where
\begin{equation}
\neigh_{{\rm out},k} \equiv \neigh_{{\rm in},k} \cup \Big( \bigcup_{\neigh_j \cap \neigh_{{\rm in},i} \neq\emptyset }  
\neigh_j \Big).
\label{newN}
\end{equation}
\end{Prop}

\noindent
{\bf Proof:} 
This is easily verified by constructing FFQLS dynamics for $\mathcal{U}(\rho)$. If $\lind_{\rm in} =\sum_i \lind_{{\rm in},i}$ is a FFQL stabilizing dynamics for $\rho_\text{in}$, construct a new Liouvillian by conjugation, that is, $\lind_\text{out} \equiv \mathcal{U} \circ\lind_\text{in} \circ\mathcal{U}^{\dagger. }$. More explicitly,
\[ \lind_\text{out} (\rho) = \prod_k U_k   \sum_i  \lind_{\text{in},i} \Big(  \prod_j U_j^\dagger \rho \prod_{j'} U_{j'}   \Big)  \prod_k U_{k'}^\dagger   
\equiv \sum_k \lind_{\text{out}, k} (\rho), \]
\noindent 
where the neighborhood 
structure of the output Liouvillian relative to the ({\em enlarged}, in general) neighborhoods in Eq. (\ref{newN}) follows from the commutativity of the circuit unitaries $U_j$, as conjugation by all but those unitaries constrained by Eq. (\ref{newN}) has no net effect. Both the spectrum and the FF property are preserved as $\mathcal{U}$ is unitary, and $\rho_\text{out}$ is stabilized by $\lind_\text{out}$ because its kernel is $\mathcal{U}(\ker(\lind_\text{in}))=\mathcal{U}(\textup{span}(\rho_\text{in}))$.
\qed

Physically, the obvious way to construct a QL commuting circuit is via exponentiation of commuting QL Hamiltonians, namely, 
${ U_j} \equiv \exp( i H_j)$, with $[H_j, H_{j'}]=0$ for all $j,j'$. In fact, {\em any} QL commuting circuit arises in this way, in the sense that a family of QL commuting Hamiltonians may always be associated to ${\cal U}$, for instance by letting $H_j = - i \log U_j$ in the basis which simultaneously diagonalizes all circuit unitaries.

\vspace*{1mm}

\noindent
{\bf Remark 5: Rapid mixing.}
As mentioned, for both pure and thermal graph states on qubits, QL stabilizing dynamics have been thoroughly analyzed in the literature. In particular, rigorous upper bounds on the {\em mixing time} have been established, showing that such states may be {\em efficiently} prepared -- that is, the (worst-case) convergence time scales only {{\em (poly-) logarithmically with the system size}}
\cite{Kastoryano2012, Temme2014}.
Remarkably, rapid mixing has been shown to both lead to stability against QL perturbations of the generator \cite{Cubitt2015} and to the emergence of effective area laws \cite{Brandao2015}. These results extend naturally to the broader classes of graph product states and encoded product FFQLS states considered here.

\subsection{Quasi-local stabilization of commuting Gibbs states} 
\label{sub:gibbs}

In this subsection, we analyze FFQLS of another class of states derived from commuting QL Hamiltonians. 
Consider a Gibbs state:
\begin{equation}
\rho_\beta \equiv \frac{e^{-\beta H}}{ \tr{}{e^{-\beta H}} }, \quad \quad H \equiv \sum_j H_j, 
\quad \beta \in {\mathbb R}^+,
\label{Gibbs}
\end{equation}
where each $H_j$ is a neighborhood-operator relative to $\neigh_j \in \neigh$. If the neighborhood Hamiltonians satisfy 
$[H_j, H_{j'}]=0$ for all $j,j',$  $\rho_\beta$ is also called a \emph{commuting Gibbs state} 
\cite{Brandao2014}.
Characterizing QL evolutions that have canonical Gibbs states as their unique fixed point has both implications for elucidating aspects of thermalization in naturally occurring dynamics and for quantum algorithms and simulation -- most notably, in the context of quantum generalizations of Metropolis sampling \cite{Temme2011}. Recent work \cite{Brandao2014} has shown that Gibbs states of arbitrary QL commuting Hamiltonians are FFQLS, the commutativity property being essential to ensure quasi-locality of either the weak-coupling (Davies) generator or the heat-bath QDS dynamics that dissipatively prepare them. The central result therein establishes an equivalence between the stabilizing dynamics being gapped and the correlations in the Gibbs state satisfying so-called ``strong clustering'', implying rapid mixing for arbitrary one-dimensional (1D) lattice systems, or for arbitrary-dimensional lattice systems at high enough temperature.
 
It is important to appreciate that in the derivation of such results, \emph{primitivity} of the QDS generator is assumed from the outset, and verified, along with the QL and FF properties, by making explicit reference to the structure of the Davies or heat-bath generator (see respectively Lemma 9 and Theorem 10 in \cite{Brandao2014}). Conversely, the QL notion is not {\em a priori} imposed as a design constraint for the dynamics, but again emerges from the structure of the generator itself. In this sense, our framework may be seen to provide a complementary approach, providing in particular a necessary condition for thermal dynamics to be primitive relative to a \emph{specified} neighborhood structure.  Let us illustrate the potential of our approach by focusing on the simplest setting of commuting two-body NN Hamiltonians in 1D. 

{\begin{Prop}
A full-rank state $\rho>0$ defined on a 1D lattice system is FFQLS relative to neighborhoods $\neigh_j=\{ j,j+1\}$ 
if and only if
\begin{equation*}
 \textup{span}(\rho) = \bigg(\bigotimes_{j\textup{ odd}} {\cal F}_{\rho_{j,j+1}}\{\Sigma_{j,j+1}(\rho)\}\bigg)\bigcap
 \bigg({\cal F}_{\rho_{1}}\{\Sigma_{1}(\rho)\}\otimes\bigotimes_{k\textup{ even}} {\cal F}_{\rho_{k,k+1}}
\{\Sigma_{k,k+1}(\rho)\}\bigg).
\end{equation*}
\end{Prop}
\vspace*{10pt}
\noindent
{\bf Proof:} 
The proof simply follows from noting that we can group the neighborhoods $\{j,j+1\}$ with odd and even $j$ and, since neighborhoods in the same group are not overlapping, that the intersection of the 
minimal fixed-point sets corresponding to neighborhoods in the same group corresponds to their product.
\qed

\vspace*{1mm}

\noindent 
One consequence of the above simplification is the following:

\begin{Prop}
Let $\rho>0$ be FFQLS with respect to the above 1D NN neighborhood structure. If, for any neighborhood $\{l,l+1\}$, 
the minimal fixed-point set is the full algebra, that is, ${\cal F}_{\rho_{l,l+1}}\{\Sigma_{l,l+1}(\rho)\}=
\mathcal{B}(\hilbert_{l,l+1})$, then $\rho$ must factor as $\rho=\rho_{1\ldots l}\otimes\rho_{l+1\ldots n}$.
\end{Prop}

\vspace*{10pt}
\noindent
{\bf Proof:} 
Assume that $\rho$ is FFQLS and ${\cal F}_{\rho_{l,l+1}}\{\Sigma_{l,l+1}(\rho)\}=\mathcal{B}(\hilbert_{l,l+1})$. 
Then, $\rho$ satisfies the intersection condition, which is simplified to a tensor product of two intersections
\begin{eqnarray}
\label{eq:intprod} \textup{span}(\rho) = (\textup{int left})_{1,\ldots, l}\otimes(\textup{int right})_{l+1,\ldots,n},
\end{eqnarray}
where $(\textup{int left})_{1,\ldots, l}$ stands for 
\begin{equation*}
\bigg[\bigg(\bigotimes_{j\textup{ odd}}^{l-2} {\cal F}_{\rho_{j,j+1}}\{\Sigma_{j,j+1}(\rho)\}\otimes\mathcal{B}(\hilbert_l)\bigg)\bigcap\bigg({\cal F}_{\rho_{1}}\{\Sigma_{1}(\rho)\}\otimes\bigotimes_{k\textup{ even}}^{l-1} {\cal F}_{\rho_{k,k+1}}\{\Sigma_{k,k+1}(\rho)\}\bigg)\bigg]
\end{equation*}
and, similarly, $(\textup{int right})_{l+1,\ldots, n}$ is given by
\begin{equation*}
\bigg[\bigg(\mathcal{B}(\hilbert_{l+1})\otimes\bigotimes_{j\textup{ odd}, j=l+2}^{n} {\cal F}_{\rho_{j,j+1}}\{\Sigma_{j,j+1}(\rho)\}\bigg)\bigcap\bigg(\bigotimes_{k\textup{ even}, k=l+1}^{n} {\cal F}_{\rho_{k,k+1}}\{\Sigma_{k,k+1}(\rho)\}\bigg)\bigg].
\end{equation*}
Eq. (\ref{eq:intprod}) can only be satisfied if $\rho=\rho_{1\ldots l}\otimes\rho_{l+1\ldots n}$.
\qed

\vspace*{1mm}}

The above proposition captures the fact that, for a state of a 1D NN-coupled chain to be FFQLS, there are limitations to the correlations that the state can exhibit. Similar restrictions are, generically, sufficiently strong to prevent Gibbs states of 1D commuting NN Hamiltonians to be FFQLS {\em relative to NN neighborhoods}. A simple example is a 1D Ising Hamiltonian:
\begin{equation}
\label{Ising}
H= \sum_{j=1}^n Z_j Z_{j+1}, \quad n>3, 
\end{equation}
where periodic boundary conditions are assumed. 
The Schmidt span for each NN pair consists of the space of diagonal matrices, and is closed under generation of the distorted algebra. The intersection of all the Schmidt spans is thus the $2^n$-dimensional space of diagonal matrices, implying that, for all temperatures and size $n$, the {\em Gibbs state is not FFQLS}. In particular, no FF thermal dynamics subject to the NN QL constraint can stabilize (or be primitive with respect to) this state.

The thermal dynamics of the Davies or heat-bath generators that stabilize commuting Gibbs states 
are, in fact, both FF and QL, albeit relative to a different neighborhood structure than the one 
solely determined by the system's Hamiltonian \cite{Brandao2014}. This 
is most transparent in the weak-coupling derivation of the QDS, whereby the evolution induced by this 
Hamiltonian, ${U}_t\equiv e^{-itH}$, effectively ``modulates'' in time the bare system-bath neighborhood coupling 
operators, in turn determining the relevant Lindblad operators in frequency space \cite{alicki-lendi}. The net effect is 
that ${U}_t$ acts as a QL commuting circuit, resulting in a neighborhood structure which is expanded with respect to 
the one associated to $H$ or to the coupling operators alone (recall Proposition \ref{QLComp}).
In the specific Ising example of Eq. (\ref{Ising}), Davies generators are QL for \emph{three-body} 
(next-to-NN, NNN for short) neighborhoods, $\neigh_j=\{ j-1, j,j+1\}$.
Similarly, one can generalize the idea and define an enlarged ``Davies QL notion''. With respect to this QL constraint, 
commuting Gibbs states may be shown to obey our necessary and sufficient condition for FFQLS, as expected on physical grounds: 

\vspace*{1mm}

\begin{Prop}
\label{prop:gibbs}
{\bf (FFQLS commuting Gibbs states)}
Gibbs states of 1D NN commuting Hamiltonians are FFQLS relative to the Davies (NNN) neighborhood structure.
\end{Prop}

\vspace*{10pt}
\noindent
{\bf Proof:} 
{\blue 
Up to normalization and letting, for convenience, $\beta H \equiv \sum_{j=1}^{n-1} H_{j, j+1}$, $0< \beta < \infty$,  
the Gibbs state of Eq. (\ref{Gibbs}) may be written as 
$\rho_\beta \equiv e^{-H_{12}}e^{-H_{23}}\ldots e^{-H_{n-1,n}}=\sigma_{12}\sigma_{23}\ldots\sigma_{n-1,n},$
where the $\sigma_{i,i+1}$ are pairwise-commuting, invertible matrices defined that are different from the identity only in NN sites. 
Our strategy is to first compute the minimal fixed-point set $\mathcal{F}_{\rho_{234}}(\Sigma_{234}(\rho))$ 
and its intersection with $\mathcal{F}_{\rho_{123}}(\Sigma_{123}(\rho))$, and then, 
by iterating, to show that the resulting intersection is $\textup{span}(\rho)$. 
To compute $\mathcal{F}_{\rho_{234}}(\Sigma_{234}(\rho))$, we first obtain the corresponding Schmidt span. Using Eq. (\ref{eq:partialinnerprod}), $\Sigma_{234}(\rho)= \textup{span}\{\tr{\overline{234}}{\rho M}\}$, for all $M\in \mathcal{B}(\hilbert_{1})\otimes\identity_{234}\otimes\mathcal{B}(\hilbert_{5,\ldots,n})$. Letting $\tau_2 \equiv \prod_{j= 5}^{n-1} \sigma_{j,j+1}$, we can write 
$$\mathcal{B}(\hilbert_{1})\otimes\identity_{234}\otimes\mathcal{B}(\hilbert_{5,\ldots,n})=\textup{span}\{\tau_2(A_{1}\otimes\identity_{234}\otimes B_{5}\otimes C_{6,\ldots,n})\},$$
where $A,B,C$ range over all matrices acting on those sites. With this parameterization, the Schmidt span is simplified to
\begin{eqnarray*}
\Sigma_{234}(\rho)& =& \textup{span}\{\tr{1}{(A_{1}\otimes \identity_{\overline 1})\sigma_{12}}\sigma_{23}\sigma_{34}\tr{5}{(B_{5}\otimes \identity_{\overline 5})\sigma_{45}}\} \\
&= &\sigma_{23}\sigma_{34}[\Sigma_2(\sigma_{12})\otimes\identity_3\otimes\Sigma_{4}(\sigma_{45})], 
\end{eqnarray*}
noting that $\sigma_{23}$ and $\sigma_{34}$ commute with all operators in this space.
To calculate $\mathcal{F}_{\rho_{234}}(\Sigma_{234}(\rho))$, we first obtain the reduced state $$\rho_{234}=\sigma_{23}\sigma_{34}\tr{\overline{234}}{\sigma_{12}\sigma_{45}\ldots\sigma_{n-1,n}}=\sigma_{23}\sigma_{34}(\sigma_{2}\otimes\identity_{3}\otimes\sigma_{4}),$$
where $\sigma_2=\tr{\overline 2}{\sigma_{12}}$ and $\sigma_4=\tr{\overline 4}{\sigma_{45}}$.
It follows that $\mathcal{F}_{\rho_{234}}(\Sigma_{234}(\rho))$ has a simple structure:
$$\mathcal{F}_{\rho_{234}}(\Sigma_{234}(\rho))=\sigma_{23}\sigma_{34}[\mathcal{F}_{\sigma_{2}}(\Sigma_{2}(\sigma_{12}))\otimes\identity_{3}\otimes\mathcal{F}_{\sigma_{4}}(\Sigma_{4}(\sigma_{45}))].$$
Direct calculation verifies that $\mathcal{F}_{\rho_{234}}(\Sigma_{234}(\rho))$ obeys the required properties of 
closure under the distortion map $\Phi_{\rho_{234}}$ and invariance under $\rho_{234}$-modular action. 
Similarly, we have
$$\mathcal{F}_{\rho_{123}}(\Sigma_{123}(\rho))=\sigma_{12}\sigma_{23}[\identity_{12}\otimes\mathcal{F}_{\sigma_{3}}(\Sigma_{3}(\sigma_{34}))].$$

Finally, we compute the intersection of these two adjacent fixed-point sets. To 
highlight the necessary structure, we write $\mathcal{F}_{\sigma_{k}}^{i,i+1}\equiv\mathcal{F}_{\sigma_{k}}(\Sigma_{k}(\sigma_{i,i+1}))$, where $k=i$ or $i+1$. The relevant intersection is then
$\{\sigma_{12}\sigma_{23}[\identity_{12}\otimes\mathcal{F}_{\sigma_{3}}^{34}\otimes\mathcal{B}(\hilbert_{4})]\}\cap\{\sigma_{23}\sigma_{34}[\mathcal{B}(\hilbert_{1})\otimes \mathcal{F}_{\sigma_{2}}^{12}\otimes\identity_{3}\otimes\mathcal{F}_{\sigma_{4}}^{45}]\}.$
Factoring out the common invertible multiple of $\sigma_{23}$, this intersection simplifies to 
\begin{eqnarray*}
&\sigma_{23}[\{(\sigma_{12})\otimes (\mathcal{F}_{\sigma_{3}}^{34}\otimes\mathcal{B}(\hilbert_{4}))\}
 \cap\{(\mathcal{B}(\hilbert_{1})\otimes \mathcal{F}_{\sigma_{2}}^{12})\otimes(\sigma_{34}\mathcal{F}_{\sigma_{4}}^{45})]\}]\\
&=\sigma_{23}[\{(\sigma_{12})\cap(\mathcal{B}(\hilbert_{1})\otimes \mathcal{F}_{\sigma_{2}}^{12})\}\otimes \{(\mathcal{F}_{\sigma_{3}}^{34}\otimes\mathcal{B}(\hilbert_{4}))\cap (\sigma_{34}\mathcal{F}_{\sigma_{4}}^{45})\}.
\end{eqnarray*}
For each intersection, notice that one argument is contained in the other, giving
\begin{eqnarray*}
\sigma_{23}[\sigma_{12}\otimes\sigma_{34}\mathcal{F}_{\sigma_{4}}^{45}]=\sigma_{12}\sigma_{23}\sigma_{34}[\identity_{123}\otimes\mathcal{F}_{\sigma_{4}}^{45}].
\end{eqnarray*}
By iterating, we find that subsequent intersections simplify to $\sigma_{12}\ldots\sigma_{j-1,j}[\identity_{1,\ldots, j-1}\otimes \mathcal{F}_{\sigma_{j}}^{j,j+1}].$ 
For the final intersection, we may take $\sigma_{n,n+1} = \identity_n\otimes 1_{n+1}$, where we take the $(n+1)$-th system to be trivial. This leads to $\mathcal{F}_{\sigma_{n}}(\Sigma_{n}(\sigma_{n,n+1}))=\textup{span}(\identity_n)$. Thus, after taking the intersection over all fixed-point sets, we are left with
\begin{equation}
\bigcap\mathcal{F}_{\rho_{j,j+1,j+2}}(\Sigma_{j,j+1,j+2}(\rho)) = 
\sigma_{12}\ldots\sigma_{n-1,n}(\identity_{1,\ldots, n-1}\otimes \text{span}(\identity_{n}))=\textup{span}(\rho),
\end{equation}
which, by Theorem \ref{thm:suffcond}, proves that these states are FFQLS.
}
\qed

\subsection{Quasi-local stabilization beyond commuting Hamiltonians}
\label{sub:nc}

So far, the identification of a commuting structure has played an important role in the verification 
of the FFQLS property. 
{Thus, it is an important question to determine the extent to which ``lack of commutativity'' 
may hinder FFQLS.  The issue is simpler and better (albeit still only partially) understood for pure target 
states, in which case families of QL stabilizable states not stemming from a commuting structure have been 
identified for arbitrary system size and complex multi-partite entanglement patterns.  
Notably, spin-$1$ AKLT states in 1D, which are the archetypal example of a valence-bond-solid state in 
condensed-matter physics \cite{AKLT}, as well as a spin-$3/2$ (or higher) AKLT states in 2D, 
which provide a resource for universal quantum computation \cite{Gavin,Andrew2016},
are unique ground states of FF anti-ferromagnet Hamiltonians.  As such, they are FFQLS using NN, 
two-body dissipative dynamics \cite{Kraus2008}.
Perhaps even more surprisingly, the FFQLS property
still holds for long-range entangled states known as {\em Motzkin states} \cite{Ramis}, 
which are also unique ground states of FF NN spin-$1$ 
Hamiltonians and have been proved to (logarithmically) {\em violate the area law}.  

In what follows, we first exhibit a family of ``non-commuting'' FFQLS multi-qudit \emph{generalized Dicke 
states}, by also including a general result linking QL stabilizability of a pure state to its ability to be uniquely 
determined by its neighborhood marginals. Focusing then on mixed target states, 
we construct and analyze two explicit (non-scalable) examples showing that commutativity of the parent 
Hamiltonian or the generating QL circuit is, as for pure states, not necessary for FFQLS in general.}

\subsubsection{Pure Dicke states on qudits}
\label{Dicke}

Since the main emphasis of this paper is on mixed states, most of the technical proofs of the results in this section 
are deferred to Appendix C.  Similar to qubit Dicke states from quantum optics \cite{Garraway2011}, 
qudit Dicke states may be constructed by symmetrizing an $n$-qudit 
product state in which $k$ of the $n$ subsystems are ``excited'' to a given single-particle state, and the 
remaining $(n-k)$ are in their ``vacuum''. 
We may further naturally generalize by allowing multi-level excitation. Specifically, 
let $\{ \ket{\ell} \}$, $\ell=0, \ldots, (d-1)$, denote an orthonormal basis in ${\mathbb C}^d$ and 
$\mathcal{S}_n \equiv \{ \pi\}$ the symmetric (permutation) group on $n$ objects. Then each generalized Dicke state is 
in one-to-one correspondence with a vector of integers, $\vec{\Lambda}=(k_0,\ldots,k_{d-1})$, 
where each $k_\ell$ specifies the occupation number (multiplicity) of each single-qudit state. That 
is\footnote{States of this form have been recently analyzed in \cite{Parashar2009}, where additionally 
non-uniform superpositions are also considered. We maintain {\em permutation symmetry} as a hallmark our generalization of standard Dicke states.}, 
\begin{equation}
\ket{\vec{\Lambda}} \equiv \ket{(\underbrace{0\ldots 0}_{k_0},\underbrace{1\ldots 1}_{k_1},\ldots,\underbrace{d-1\ldots d-1}_{k_{d-1}})}\equiv\frac{1}{n!}\hspace*{-1mm}
\sum_{\pi\in\mathcal{S}_n}V_{\pi}\ket{\underbrace{0\ldots 0}_{k_0}\underbrace{1\ldots 1}_{k_1}\ldots\underbrace{d-1 \ldots d-1}_{k_{d-1}}} , 
\label{DickeS}
\end{equation}
where $\sum_i k_i = N$ and $V_{\pi}$ permutes the subsystems according to the permutation $\pi$. 
A useful fact to our purpose is that generalized Dicke states admit a simple Schmidt decomposition. Consider a partition of the system into two groups of $n_A$ and $n_B$ subsystems, respectively. It is then easy to show that the Schmidt decomposition of $\ket{\vec{\Lambda}}$ is
\begin{equation}
\ket{\vec{\Lambda}}=\frac{1}{\sqrt{{n \choose {\vec{\Lambda}}}}}\sum_{\vec{\Lambda}_A+\vec{\Lambda}_B=\vec{\Lambda}}\mu_{\vec{\Lambda}_A,\vec{\Lambda}_B}\ket{\vec{\Lambda}_A}\otimes\ket{\vec{\Lambda}_B},
\quad \mu_{\vec{\Lambda}_A,\vec{\Lambda}_B}=
\sqrt{{n_A \choose \vec{\Lambda}_A}{n_B \choose \vec{\Lambda}_B}}.
\label{DickeSch}
\end{equation}

{In order to specify the relevant class of FFQLS generalized Dicke states, 
the choice of the neighborhood structure is crucial. The following definition captures the required feature:}

\begin{defin}
{A neighborhood structure $\{\neigh_k\}$ is \emph{connected} if for any bipartition of the subsystems, 
there is some neighborhood containing subsystems from both parts.}
\end{defin}

\vspace*{1mm}

\noindent
Our main result is then contained in the following:

\begin{Prop}
{\bf (FFQLS Dicke states})
Given $n$ qudits and a connected neighborhood structure $\neigh$, there exists a (non-factorized) FFQLS 
generalized Dicke state relative to $\neigh$ if $d(m -1) \geq n$, where $m$ is the size of the 
{\em largest} neighborhood in $\neigh$.
\label{DickeQL}
\end{Prop}

\noindent 
The proof (given in Appendix C) is constructive, and yields in particular the state 
\begin{equation}
\label{eq:bigdicke}
D_{n,m}\equiv\ket{(\underbrace{0\ldots 0}_{m-1},\underbrace{1\ldots 1}_{m-1},\ldots,
\underbrace{d-1\ldots d-1}_{r})}, \quad r=n-(d-1)(m-1), 
\end{equation}
as a non-factorized (entangled) DQLS state for given, arbitrary system size.
Nonetheless, note that the product of the neighborhood size and qudit dimension must be scaled accordingly. 
For example, if $\neigh$ is fixed to be two-body NN (hence $m=2$), then the qudit dimension itself must be at least 
$d=n$. In this sense, the resulting family of states is non-scalable. 

\vspace*{1mm}
\noindent
{\bf Remark 6.}
As a particular case of Propositions \ref{DickeQL}, we recover the fact (established in \cite{ticozzi-ql}) that the state 
$\ket{(0011)}= D_{4,3}$,
also previously defined in Eq. (\ref{D4}), is FFQLS with respect to the 
neighborhood structure $\neigh_1=\{1,2,3\}$, $\neigh_2=\{2,3,4\}$. 

\vspace*{1mm}

Generalized Dicke states are non-trivially entangled. Their multiparty correlations have the feature of being 
\emph{uniquely} determined by a proper subset of all possible marginals \cite{Parashar2009}, that is, of being 
``uniquely joined'' \cite{JohnsonPRA}. 
Remarkably, an interesting connection may be made between the extent to which an 
{\em arbitrary pure} target state is uniquely determined by the set of its neighborhood-marginals
and the FFQLS property. This is formalized in the following:

\begin{Prop}
\label{unique}
{\bf (Unique joinability)}
If a pure state $\ket{\psi}$ is FFQLS relative to $\neigh$, then $\ket{\psi}$ is {\em uniquely} 
determined by its neighborhood reduced
states $\{\rho_{\neigh_j}\} = \{ \,\text{Tr}_{\overline{\neigh}_j} ( {\ket{\psi}\bra{\psi}}) \}   $. 
\end{Prop}

\vspace*{10pt}
\noindent
{\bf Proof:} 
By contradiction, if $\ket{\psi}$ is not uniquely joinable, then there exists a state {$\tau$, not necessarily pure, 
such that $\tau \ne \ket{\psi}\bra{\psi}$ and 
$\text{Tr}_{\overline{\neigh}_j} (\tau) = \rho_{\neigh_j}$, for all $\neigh_j\in \neigh$. }Clearly, $\supp(\tau)
\leq  \supp(\rho_{\neigh_j}) \otimes \mathcal{B}(\Hi_{\overline{\neigh}_j})$, for all $j$, hence by recalling Theorem 
\ref{mainthm} it also follows that 
\[ \supp(\tau) \leq \bigcap_j \supp(\rho_{\neigh_j}) \otimes \mathcal{B}(\Hi_{\overline{\neigh}_j}) = \span(\ket{\psi}). \]
On the other hand, since by assumption $\supp(\tau) \ne \span(\ket{\psi})$, the above inclusion must be strict, which is 
impossible since $\span(\ket{\psi})$ is one-dimensional. 
\qed

\vspace*{1mm}

{\blue
We now establish that the class of states of Eq. (\ref{eq:bigdicke}) constitute genuinely non-commuting examples of FFQLS. 
To do so, we show that, with respect to a certain class of neighborhood structures, these states are the unique ground states 
of non-commuting FF QL Hamiltonians, but they {\em cannot} be the unique ground states of {\em any} 
commuting FF QL Hamiltonians.

First, since the ground-state space of a FF Hamiltonian is simply the intersection of the ground-state spaces of the 
individual Hamiltonian terms in the sum, replacing each such term with the projector onto its excited space preserves 
the ground space of the global FF Hamiltonian.  Without loss of generality, we may then restrict attention to QL 
Hamiltonians consisting of sums of  neighborhood-acting projectors.
An important consequence of this simplification is that for a given FFQL Hamiltonian, the only possible candidates for other FFQL Hamiltonians with the {\em same} ground-state space are sums of projectors with enlarged ground-state spaces, with respect to those of the given Hamiltonian.
However, the following Lemma demonstrates that, in the two-neighborhood case, a non-commuting FFQL Hamiltonian does not admit a commuting enlargement with the same ground state space:

\begin{lemma} 
{\bf (Commuting enlargements)}
\label{thm:enlarged}
Assume that $\Pi^0_1,\Pi_2^0$ are two non-commuting 
projections such that
\[\lim_{n\rightarrow\infty}(\Pi^0_1\,\Pi_2^0)^n \equiv \Pi_{\rm{GS}}^0\neq 0. \]
If $\,\Pi_1,\Pi_2$ are enlarged commuting projectors such that 
\beq 
\Pi_{k}\,\Pi_{k}^0=\Pi_{k}^0\,\Pi_{k}=\Pi_{k}^0, \quad k=1,2, 
\label{eq:enlarged}
\eeq
then $\Pi_{1}\Pi_{2}\equiv \Pi_{\rm{GS}}$, with $\Pi_{GS}\gneq \Pi_{\rm{GS}}^0.$
\end{lemma}

\vspace*{10pt}
\noindent
{\bf Proof:} 
Using Eq. (\ref{eq:enlarged}), we have that 
\[ \Pi^0_1\,\Pi_2^0 = \Pi^0_1\, \Pi_1\,\Pi_2\,  \Pi_2^0,\quad  \quad \Pi^0_2\,\Pi_1^0 = \Pi^0_2\, \Pi_2\,\Pi_1\,  \Pi_1^0. \]
\noindent 
Towards proof
by contradiction, assume that $\Pi_{\rm{GS}}=\Pi_1\Pi_2 = \Pi_{\rm{GS}}^0$. It follows that
\[ \Pi^0_1\,\Pi_2^0 = \Pi_1^0\,  \Pi^0_{\rm{GS}} \,\Pi^0_2 , \quad\quad  \Pi^0_2\,\Pi_1^0 = \Pi_2^0\, \Pi^0_{\rm{GS}}\, \Pi_1^0.\]
\noindent 
By using the defining property of $\Pi^0_1,\Pi_2^0$, however, the right hand-side in each of the above equalities simplifies to 
$\Pi_k^0\, \Pi^0_{\rm{GS}}\, \Pi_k^0= \Pi^0_{\rm{GS}}$, for 
$k=1,2$. This in turn yields 
$$ \Pi_1^0\, \Pi_2^0 = \Pi^0_{\rm{GS}} = \Pi^2_0 \, \Pi_0^1, $$ 
which contradicts the non-commuting assumption.
\qed

\vspace*{1mm} 

With this Lemma in place, we now verify the genuine non-commutativity of these states.

\begin{Prop}
{\bf (Non-commutativity of FFQLS Dicke states)}
For each (non-factorized) Dicke state $D_{n,m}$ of Eq. (\ref{eq:bigdicke}), there exists a neighborhood structure for 
which $D_{n,m}$ cannot be the unique ground state of any commuting FF QL Hamiltonian.
\end{Prop}
\vspace*{10pt}
\noindent
{\bf Proof:} 
Consider the (non-factorized) Dicke state $D_{n,m}$. Correspondingly, we choose any connected neighborhood structure $\neigh$ with $m$-body neighborhoods, whereby there is {\em at least one} neighborhood ($\neigh_1$, say) containing a system that is not contained in any other neighborhood. 
Such a neighborhood structure always exists, and Proposition \ref{DickeQL} ensures that $D_{n,m}$ is FFQLS relative to that. 
From \cite{ticozzi-ql}, any FFQLS state $\ket{\psi}$ is the unique ground state of some FF QL Hamiltonian.
In particular, letting $\ket{\psi} \equiv D_{n,m}$, one such parent Hamiltonian is  
\begin{equation}
H= \sum_j H_j \equiv \sum_j ( \identity-\Pi_{\neigh_j}\otimes\identity_{\overline{\neigh}_j}),
\label{nc}
\end{equation}
where $\Pi_{\neigh_j}$ is the projector onto the Schmidt span of 
$D_{n,m}$ with respect to $\neigh_j$. 
We first show that the $H_j$ do not commute with one another. From the Schmidt decomposition of $D_{n,m}$, 
given in Eq. (\ref{DickeSch}), we have $\Pi_{\neigh_j}=\sum_{\Lambda}\ketbra{\Lambda}$, where the sum extends 
over all choices of $m$ symbols from the symbols in $D_{n,m}=\ket{\vec{\Lambda}}$. A direct calculation the shows
that the $\Pi_{\neigh_j}$, and therefore, the $H_{j}$, of overlapping neighborhoods do not commute with one another.

In order to establish the desired result, note that enlarging the neighborhoods preserves FFQLS. 
Hence, $D_{n,m}$ is also FFQLS with respect to the two-neighborhood neighborhood structure with one 
neighborhood being $\neigh_1$ and the other neighborhood, say, $\neigh_U$, being the union of the remaining neighborhoods 
of $\neigh$. Furthermore, by building a parent Hamiltonian out of these two projectors as in Eq. (\ref{nc}), the reasoning above 
shows that the projectors of $D_{n,m}$ with respect to $\neigh_1$ and $\neigh_U$ do not commute. Hence, by Lemma \ref{thm:enlarged}, no commuting enlargement exists for $\{\neigh_1$, $\neigh_U\}$. Assume now 
that a commuting 
QL parent Hamiltonian exists for the {\em original} neighborhood structure. The projectors $\Pi_k$ onto the ground state 
spaces of these Hamiltonians also commute. Along with FF condition, this implies that
$$  \ketbra{\psi}= \Pi_1 \Pi_2 \ldots \Pi_{|\neigh|} \equiv \Pi_1 \Pi_U,$$
where $\Pi_U \equiv \Pi_2 \ldots \Pi_{|\neigh|}$. Hence, the QL Hamiltonian $H= (\identity-\Pi_1) + (\identity-\Pi_U)$ constitutes a commuting enlargement for $\{\neigh_1$, $\neigh_U\}$, which we showed cannot exist. This contradiction then implies that no commuting enlargement can exist for $\neigh$.
\qed
}

\subsubsection{Non-commuting Gibbs states} 

In order to demonstrate that genuinely mixed target states may also be FFQLS despite not being obviously associated 
to a commuting structure, specific examples may be constructed in 1D by considering a generalization of the NN Ising 
Hamiltonian considered in Eq. (\ref{Ising}), obtained by adding a transverse (magnetic) field. That is\footnote{Hamiltonians such as in Eq. (\ref{trans}) are exactly solvable upon mapping to a free-fermion problem, and {\emph{commuting} once expressed in terms of appropriate quasi-particles. }Indeed, thermal dynamics associated to free-fermion QDS is known to be hypercontractive \cite{Temme2014},  {\em in the absence} of a QL ``real-space'' constraint as we consider.}:
\begin{equation}  
H= - \sum_{j=1}^{n-1} Z_j Z_{j+1} - g \sum_{j=1}^n X_j,  
\quad \quad n \geq 4, \quad g \in {\mathbb R}^+ .
\label{trans}
\end{equation}
\noindent 
In particular, we consider (full-rank) Gibbs states, constructed as in Eq. (\ref{Gibbs}), as well as variants inspired 
by non-equilibrium quench protocols, wherein an initial thermal state of a Hamiltonian with given $g$ is evolved under a Hamiltonian with $g'\ne g$. In all cases, the relevant minimal distorted algebras and their intersection have been numerically constructed (in Matlab) for given QL constraints, and Theorem \ref{thm:suffcond} used to determine FFQLS. 
The results are found to depend sensitively on the neighborhood structure: Gibbs (and generalized Gibbs) states 
on $n=4$ qubits are found to be FFQLS for three-body neighborhoods (as in the corresponding commuting Ising 
case), however extending to $n=5,6$ qubits requires neighborhoods to be further enlarged to allow for four-body 
Liouvillians. 

While a direct (numerical) verification is beyond reach, this points to the possibility that the (maximal) neighborhood 
size will have to scale extensively as $n$ increases, thereby preventing scalable FFQLS.
An intuitive argument in support of this is the observation that, 
as the size of the ``neighborhood complements'' increase, the dimensions of the extended Schmidt spans do as well; correspondingly, the uniqueness of their intersection, as is required for FFQLS, becomes less likely.
Despite this limitation, these results show the general applicability of our framework.

{\subsubsection{Entangled mixed states} }
\label{sec:suppcounter}

{As a final application, we analyze QL stabilizability of a one-parameter family of mixed entangled states 
on $n=4$ qubits. Beside illustrating the full procedure needed to check if $\rho$ is FFQLS and to construct the stabilizing maps, this example is useful for a number of reasons: 
first, it reinforces that genuinely multipartite entangled mixed states {\em can} be FFQLS; second, it explicitly shows that the support condition under which we proved sufficiency for general target states in Theorem \ref{thm:suffcondnorank} is 
\emph{not} necessary in general; 
lastly, it shows how a non-FFQLS state may still admit arbitrarily close (in Hilbert-Schmidt space) states that are FFQLS, that is, in control-theoretic language, it may still in principle allow for ``practical stabilization''. 
The family of mixed states we analyze may be parametrized as follows:
\begin{equation}
\rho_\epsilon \equiv (1-\epsilon) \,\ketbra{\textup{(0011)}}+ \epsilon \, \ketbra{\textup{GHZ}^4_2}, \quad \epsilon 
\in (0,1),  
\label{4Q}
\end{equation}
with neighborhoods $\neigh_1 = \{1,2,3\},$ $\neigh_2=\{2,3,4\}$. }
Here, $\textup{GHZ}^4_2$ is the usual GHZ state on qubits, that is, $\ket{\textup{GHZ}^n_d}
=\left(\ket{0}^{\otimes n}+\ldots+\ket{d-1}^{\otimes n}\right)/\sqrt{d}$. As established in \cite{ticozzi-ql}, 
this state is {\em not} DQLS for any non-trivial neighborhood structure\footnote{
GHZ states are graph states, though with respect to a \emph{star graph} (i.e., a central node connected to $(n-1)$ surrounding nodes). The neighborhood structure induced by this graph is trivial, in that it consists of a single neighborhood containing all $n$ qubits. Hence, we treat GHZ states as separate from graph states, considering 
only non-trivial QL constraints.}, 
as one may verify by seeing that the $d$-dimensional space 
$\textup{span}\{\ket{0,0,\ldots, 0},\ldots,\ket{(d-1,d-1,\ldots, d-1}\}$ is contained in each extended Schmidt span, 
and hence their intersection is greater than just $\textup{span}(\ket{\textup{GHZ}^n_d})$.

Let us use the notation $123|4$ to denote the partition of the index set $\{1,2,3,4\}$ in the neighborhood $\{1,2,3\}$ and the remaining index $\{4\},$ and similarly for $1|234$: the {\em $1|234$-Schmidt decomposition} means the Schmidt decomposition with respect to such bipartition. In order to construct QL FF dynamics which render $\rho_\epsilon$ GAS, the first step is to 
compute the operator Schmidt span for each neighborhood.
Two properties aid our analysis. First, both the Dicke and GHZ components are permutation symmetric, so that the analysis of the $1|234$ partition carries over to that of $123|4$. 
Second, they have compatible Schmidt decompositions, in the sense that we can find a {\em single} operator basis in which to Schmidt-decompose both of them and their mixtures. Using that
\begin{eqnarray*}
 \ket{\textup{(0011)}} &=&\sqrt{1/2}\ket{0}\ket{(011)}+\sqrt{1/2}\ket{1}\ket{(001)},\\
 \ket{\textup{GHZ}^4_2} &=& \sqrt{1/2}\ket{0}\ket{000}+\sqrt{1/2}\ket{1}\ket{111},
\end{eqnarray*}
the desired operator Schmidt decomposition is
\begin{eqnarray*}
 \rho_\epsilon &=& \frac{1}{2}\ketbra{0}\otimes [(1-\epsilon) \ketbra{(011)}+ \epsilon \ketbra{000}] \\
&+& \frac{1}{2}\ket{0}\bra{1}\otimes[ (1-\epsilon) \ket{(011)}\bra{(001)}+ \epsilon \ket{000}\bra{111}]\\
&+& \frac{1}{2}\ket{1}\bra{0}\otimes[ (1-\epsilon) \ket{(001)}\bra{(011)}+ \epsilon \ket{111}\bra{000}]\\
&+& \frac{1}{2}\ketbra{1}\otimes[(1-\epsilon) \ketbra{(001)}+ \epsilon \ketbra{111}].
\end{eqnarray*}
Let us focus on the factors relative to subsystems $\{234\}$ from each term above. 
{To compute the minimal fixed-point set containing this Schmidt span}, we first undo the distortion of the 
elements of this space by conjugating with respect to
\begin{equation*}
 \rho^{-\frac{1}{2}}_{234}=\frac{1}{\sqrt{1-\epsilon}}(\ketbra{(011)}+\ketbra{(001)})+\frac{1}{\sqrt{\epsilon}}(\ketbra{000}+\ketbra{111}),
\end{equation*}
where, as noted, the above inverses are taken as the Moore-Penrose inverse (with $\frac{1}{\sqrt{\epsilon}}$ or $\frac{1}{\sqrt{1-\epsilon}}$ being replaced by 0 in the singular cases of $\epsilon=0$ or $1$, respectively). 
{Conjugation of each Schmidt basis element with respect to this $\rho^{-\frac{1}{2}}_{234}$ removes the 
$\epsilon$-dependence, namely, 
\begin{eqnarray*}
&&\ketbra{(011)}+\ketbra{000}, \quad \quad \ket{(011)}\bra{(001)}+\ket{000}\bra{111},\\
&&\ket{(001)}\bra{(011)}+\ket{111}\bra{000},\quad \quad \ketbra{(001)}+\ketbra{111}.
\end{eqnarray*}}
\noindent Via a unitary change of basis, 
we identify computational basis elements $\ket{\tilde{0}\tilde{0}\tilde{0}},\ket{\tilde{0}\tilde{0}\tilde{1}},$ etc., with vectors
 \begin{equation}
\label{basisnew}
\{\ket{000}, \ket{(011)},\ket{111},\ket{(001)}, \ket{e_1}, \ket{e_2},\ket{e_3},\ket{e_4}\},
\end{equation}
where $\ket{e_1}, \ket{e_2},\ket{e_3},\ket{e_4}$ are chosen to ensure orthonormality. 
This transformation reveals that the Schmidt-span operators share a common identity factor, as in this basis they read:
\begin{equation*}
(\ketbra{\tilde{0}}\otimes \identity)\oplus {\mathbb O},\,\,(\ket{\tilde{0}}\bra{\tilde{1}}\otimes \identity)\oplus {\mathbb O},\,\,(\ket{\tilde{1}}\bra{\tilde{0}}\otimes \identity)\oplus {\mathbb O},\,\,(\ketbra{\tilde{1}}\otimes \identity)\oplus {\mathbb O},
\end{equation*}
where the sector on which the zeros act is $\textup{span}\{\ket{e_1},\ldots,\ket{e_4}\}$.
The span of these operators is closed under $*$-algebra operations and constitutes a representation of the 
Pauli algebra. Thus, the distorted Schmidt spans of $\rho_\epsilon$ are already *-closed algebras. 
{A simple calculation verifies that each distorted-algebra basis element is also unchanged by 
${\mathcal M}_{\frac{1}{2}}$. }
To find the minimal fixed-point sets, we need to apply the distortion map again, by conjugating 
the generators with $\rho^{\frac{1}{2}}_{234}.$
We can write the Schmidt decomposition, with respect to the basis in Eq. \eqref{basisnew}, as
\begin{equation*}
\ketbra{\tilde{0}}\otimes \tau\oplus {\mathbb O},\,\,\ket{\tilde{0}}\bra{\tilde{1}}\otimes \tau\oplus {\mathbb O},\,\,\ket{\tilde{1}}\bra{\tilde{0}}\otimes \tau\oplus {\mathbb O},\,\, \ketbra{\tilde{1}}\otimes \tau\oplus {\mathbb O},
\end{equation*}
where we have defined $\tau \equiv \epsilon \ketbra{\tilde{0}}+ (1-\epsilon) \ketbra{\tilde{1}}$.

The last step is to construct QL Liouvillians for each neighborhood. The requirement is that the 
kernels of each of these {are the corresponding minimal fixed-point sets}. This can be obtained by 
considering the operators:
\begin{eqnarray*}
{L}_0 = \ket{\tilde{0}}\bra{\tilde{1}}\otimes\identity\otimes\identity , \quad 
{L}_+ = \ket{\tilde{0}}\bra{\tilde{0}}\otimes\identity\otimes {\tau^\frac{1}{2}}\, \ket{\tilde{0}}\bra{\tilde{1}}, 
\quad 
L_- = \ket{\tilde{0}}\bra{\tilde{0}}\otimes\identity\otimes {\tau^\frac{1}{2}}\, \ket{\tilde{1}}\bra{\tilde{0}}.
\end{eqnarray*}
{The first Lindblad operator $L_0$ is responsible for asymptotically preparing the subspace\\ 
$\textup{span}\{\ket{000}, \ket{(011)},\ket{111},\ket{(001)}\}$, while $L_+$ and $L_-$ stabilize the $\tau$ factor. 
All three Lindblad operators must commute with the distorted algebra in order that it be preserved.}
Using the standard definitions of ladder operators, $\sigma^{+}\equiv \ket{0}\bra{1}=(\sigma^{-})^{\dagger}$, 
we rewrite the $\tau$-preparing Lindblad operators back in the original basis, in terms of standard Pauli 
matrices, as
\begin{eqnarray*}
L_+  =& \sqrt{\epsilon} & [\,\sigma^+_2\sigma^+_3\sigma^+_4\,(\sigma^-_2+\sigma^-_3+\sigma^-_4)+ 
\sigma^-_2\sigma^-_3\sigma^-_4 \,(\sigma^+_2+\sigma^+_3+\sigma^+_4)],\\
L_-  =& \sqrt{1-\epsilon} & [(\sigma^-_2+\sigma^-_3+\sigma^-_4)\,\sigma^+_2\sigma^+_3\sigma^+_4 + 
(\sigma^+_2+\sigma^+_3+\sigma^+_4)\,\sigma^-_2\sigma^-_3\sigma^-_4],
\end{eqnarray*}
where now $L_-=\sqrt{\frac{1-\epsilon }{\epsilon}}\, L^{\dagger}_+.$ 
Defining $\ket{(001)_{\omega}}\equiv (\ket{001}+\omega\ket{010}+\omega^2\ket{100})/\sqrt{3}$, 
$\omega\equiv e^{2\pi i/3}$, 
and similar terms to denote symmetric basis elements for the four dimensional space orthogonal to the symmetric 
subspace, the third Lindblad operator reads
\begin{eqnarray*}
L_0 = \ket{000}\bra{(001)_{\omega}}+\ket{(011)}\bra{(001)_{\omega^2}}
+\ket{111}\bra{(011)_{\omega}}+\ket{(001)}\bra{(011)_{\omega^2}}.
\end{eqnarray*}
\noindent 
These Lindblad operators form the neighborhood Liouvillian on systems $234$, namely,
\begin{equation*}
 \lind_{234}(\rho)=L_0\rho L_0^{\dagger}+L_+\rho L_+^{\dagger}+L_-\rho L_-^{\dagger}-\frac{1}{2}\{ L_0^{\dagger}L_0+L_+^{\dagger}L_++ L_-^{\dagger}L_-, \,\rho \},
\end{equation*}
The global generator $\lind$ is obtained by constructing $\lind_{123}$ in an analogous way, and by 
letting $\lind = \lind_{234}+\lind_{123}.$

{Using Matlab, we have verified that these dynamics are FF and stabilize $\rho_\epsilon$; 
the kernel of $\lind$ is equal to $\textup{span}(\rho_\epsilon)$, as desired. 
In the limiting cases of $\epsilon=0,1$, the above dynamics fail to have a unique fixed state. 
As we have already established, for the Dicke-state case of $\epsilon=0$, FFQL stabilizing dynamics 
can be constructed by a separate procedure, whereas for the GHZ case of $\epsilon=1$, 
no FFQL stabilizing dynamics exists relative to the given neighborhoods. }

\vspace*{1mm}

\noindent
{\bf Remark 7: Failure of support condition.}
In Theorem \ref{thm:suffcondnorank}, {in order to obtain a general sufficient condition for FFQLS},
we supplemented the necessary condition with the ``support condition'' of Eq. (\ref{eq:suppcond}). 
We now show that $\rho_\epsilon$ fails the support condition despite being FFQLS. 
The support of $\rho_\epsilon$ is spanned by just $\ket{(0011)}$ and $\ket{\textup{GHZ}^4_2}$. 
On the other hand, we also have
\begin{eqnarray*}
 \textup{supp}(\rho_{123}\otimes\identity_4)&=&\textup{span}\{\ket{0000},\ket{0001},\ket{(001)}\ket{0},\ket{(001)}\ket{1},\\
&&\ket{(011)}\ket{0},\ket{(011)}\ket{1},\ket{1110},\ket{1111} \},
\end{eqnarray*}
\vspace*{-8mm}  
\begin{eqnarray*}
 \textup{supp}(\identity_1\otimes\rho_{234})&=&\textup{span}\{\ket{0000},\ket{1000},\ket{0}\ket{(001)},\ket{1}\ket{(001)},\\
&&\ket{0}\ket{(011)},\ket{1}\ket{(011)},\ket{0111},\ket{1111}\}.
\end{eqnarray*}
The intersection is found to be
\begin{equation*}
 \textup{span}\{\ket{0000},\ket{(0001)},\ket{(0011)},\ket{(0111)},\ket{1111}\}.
\end{equation*}
Since this intersection properly contains the support of $\rho_\epsilon$, the claim follows. 

\vspace*{1mm}

\noindent 
{\bf Remark 8.}
{Since stabilization of $\rho_\epsilon$ is possible for $\epsilon$ \emph{arbitrarily close} to the GHZ-value of one, 
the above example demonstrates that there exist
sequences of FFQLS mixed states that converge to a non-FFQLS pure state.
While this allows for practical stabilization \cite{khalil-nonlinear} in principle, we expect that the 
Liouvillian spectral gap will close as the non-FFQLS state is approached -- making stabilization inefficient.
{If we normalize the Liouviallians to $||\lind||_2=1$, the corresponding gaps 
$\Delta$ are found to behave as $\Delta \approx 0.049 \,(1-\epsilon)$.}
It remains an interesting question for future investigation to determine whether similar conclusions about QL 
practical stabilization and associated efficiency trade-offs may be drawn for more general 
target states. }

\vspace*{4mm}

\section{Outlook}
\label{sec:end}

We have introduced a systematic method, following a system-theoretic approach, to determine whether a 
general {\em mixed} state of a finite-dimensional multi-partite quantum system may be the unique 
fixed point for a natural class of QL FF Markovian dynamics -- for {\em given} locality constraints. 
If appropriate conditions are obeyed, a constructive procedure to synthesize stabilizing  
dynamics is provided. For states that are not full-rank, the same necessary conditions hold, however 
a method for constructing stabilizing generators is provided only if an additional ``support'' condition is 
satisfied. In both cases, stabilization may be achieved without requiring Hamiltonian control resources 
in principle. We have presented a number of QIP and physically motivated examples demonstrating how our tools can 
naturally complement and genuinely extend available techniques for fixed-point convergence and 
stability analysis -- including QL stabilization of Gibbs states of {\em non-commuting} Hamiltonians
(albeit for finite system size).  Altogether, beside filling a major gap in the existing pure-state QL stabilizability analysis,
we believe that our results will have direct relevance to dissipative QIP and quantum engineering, 
notably, open-system quantum simulators. 

A number of research questions are prompted by the present analysis and call 
for further investigation. Determining whether our necessary condition for FFQLS is, as we  
conjecture, \emph{always} sufficient on its own even for non-full-rank states is a first obvious issue
to address.  From a physical standpoint, in order to both understand the role of Hamiltonan control 
and to make contact with naturally occurring dissipative dynamics, it is important to scrutinize 
the extent to which the stabilizing dynamics obtained in our framework may be compatible with 
rigorous derivations of the QDS -- in particular in the weak-coupling-limit, where the interplay between Hamiltonian 
and dissipative components is crucial and demands to be carefully accounted for \cite{alicki-lendi,Domenico}.  
Still within the present QL stabilization setting, an interesting mathematical question 
is to obtain a global characterization of the {\em geometrical and topological properties} 
of the FFQLS set, beginning from pure states. As alluded to in the last remark above, answering these 
questions may also have practical implications, in terms of {\em approximate} QL stabilization.
Related to that, while our main focus has indeed been on {\em exact} asymptotic stabilization, 
it may also be beneficial (possibly necessary) to tailor approximate methods for analysis and/or 
synthesis to specific classes of states. A natural starting point could be provided here by  
graph states, which have recently been shown to arise as {\em arbitrarily accurate}  
approximations of ground states of {\em two-body} FF Hamiltonians \cite{Darmawan}. 
  
From a more general perspective, as remarked in the Introduction, the present analysis 
fits within our broader program of understanding {\em controlled open-quantum system dynamics 
subject to a resource constraint}. In that respect, an important next step will be to tackle 
different kinds of constraints -- including, for instance, less restrictive notions of 
quasi-locality, which allow for exponentially decaying interactions in space (as in 
\cite{Brandao2014}); or possibly reformulating the QL constraint away from ``real'' space and 
the associated tensor-product-decomposition, but rather relative to 
a preferred operator subspace, in the spirit of ``generalized entanglement'' \cite{Barnum2003}. 
Lastly, and perhaps somewhat counter-intuitively, as our results on separable non-FFQLS 
evidence, it is interesting to acknowledge that the QL stabilization problem need not be trivial for 
{\em classical} probability distributions. For both the classical and the quantum setting, we find
it tempting to speculate that interesting mathematical connections may exist with the 
quantum marginal problem and general quantum joinability \cite{JohnsonPRA,JohnsonJPA}.

\section{Acknowledgements}
\noindent
We thank Fernando Brandao for valuable input, 
and Veronica Umanit\`a and Emanuela Sasso for pointing out an issue with a previous version of a proof. 
Work at Dartmouth was partially supported by the National Science Foundation under grant No.
PHY-1104403, the Army Research Office under contract No. W911NF-14-1-0682, 
and the Constance and Walter Burke {\em Special Projects Fund in Quantum 
Information Science}.  F.T. has been partially funded by the QUINTET, QFUTURE and QCOS 
projects of Universit\`a degli Studi di Padova. P.D.J. also gratefully acknowledges support 
from a {\em Gordon F. Hull} Dartmouth graduate fellowship.

\section*{Appendix A: Equivalence between FFQLS and DQLS for pure fixed points}

\noindent 
{\bf Proposition A.1.}
{\bf (Pure-state FFQLS is DQLS)}
{\em  A pure state $\rho=\ketbra{\psi}$ is DQLS if and only if it is FFQLS.}

\vspace*{10pt}
\noindent
{\bf Proof:} 
If $\ketbra{\psi}$ is DQLS, then by definition there exists QL dynamics with no Hamiltonian 
component and QL noise operators satisfying $D_k\ket{\psi} = 0$ in standard form. Then, clearly, both 
$\mathcal{L}(\ketbra{\psi})=0$ and $\mathcal{L}_{\neigh_j}(\ketbra{\psi})=0$ for each neighborhood, 
making $\ket{\psi}$ FFQLS.

To prove the converse implication, note that it follows from Theorem \ref{mainthm} (proved in \cite{ticozzi-ql})
that a pure state $\ketbra{\psi}$ is DQLS if and only if 
$\textup{span}(\ketbra{\psi})=\bigcap_j \mathcal{B}(\Sigma_{\neigh_j}(\ket{\psi})\otimes
\hilbert_{\overline{\neigh}_j}). $
We show that this equality follows if $\ketbra{\psi}$ is FFQLS. If so, 
then Theorem \ref{thm:neccond} implies that 
{
$$  \textup{span}(\ketbra{\psi})=\bigcap_j {\cal F}_{\rho_{\neigh_j}}(\Sigma_{\neigh_j}(\ketbra{\psi}))
 \otimes\mathcal{B}(\hilbert_{\overline{\neigh}_j}) $$ 
The Schmidt span of a pure density operator $\ketbra{\psi}$ is simply $\Sigma_{\neigh_j}(\ketbra{\psi})=\mathcal{B}[\textup{supp}(\rho_{\neigh_j})]$. Then, $\mathcal{B}[\textup{supp}(\rho_{\neigh_j})]$ is already a 
modular-invariant $\rho_{\neigh_j}$-distorted algebra, thus  
we do not need to enlarge it to obtain ${\cal F}{\rho_{\neigh_j}}(\Sigma_{\neigh_j}(\ketbra{\psi}))\otimes\mathcal{B}(\hilbert_{\bar{\neigh}_j}).$ }
With this, and the fact that $\mathcal{B}(V)\otimes\mathcal{B}(W)=\mathcal{B}(V\otimes W)$, we may write
$$\textup{span}(\ketbra{\psi})=\bigcap_j\mathcal{B}(\textup{supp}(\rho_{\neigh_j}\otimes\identity_{\bar{\neigh}_j})).
$$ 
Lastly, we use the properties that $\bigcap_i \mathcal{B}(V_i)=\mathcal{B}(\bigcap_i V_i)$ and 
$\mathcal{B}(V)=\mathcal{B}(W)\Leftrightarrow V=W$ to reduce the above equality from operator 
space to vector space,
$$ \textup{supp}(\ketbra{\psi})=\bigcap_j \textup{supp}(\rho_{\neigh_j}\otimes\identity_{\bar{\neigh}_j}), $$
implying that the target state $\ket{\psi}$ is DQLS, as desired. 
\qed

This result yields a characterization of FF dynamics which uniquely stabilize a pure state:

\vspace*{2mm}

\par\noindent 
{\bf Corollary A.2.}
{\em If a pure state $\rho=\ketbra{\psi}$ is FFQLS, then there exists a generator that makes $\ketbra{\psi}$ FFQLS 
and is of purely-dissipative form:
$$ \mathcal{L}(\cdot)=\sum_k D_k \cdot D_k^{\dagger} -\frac{1}{2}\{D_k^{\dagger}D_k,\cdot\},$$
where each Lindblad operator $D_k$ is a neighborhood operator 
and, in addition, obeys $D_k\ket{\psi}=0$ for each $k$.}

\section*{Appendix B: A result on fixed states of convex combinations of CPTP maps}
\label{app:keyresult-general}

\noindent 
{\bf Theorem B.1.}
{\bf (Common fixed points of sums of CPTP maps)}
{\em Let $\Ti=\sum_{k} p_k\Ti_{k}$ be a sum of CPTP maps, with $p_k>0$ and $\sum_k p_k=1$, and 
assume that both the following conditions hold: \\
\noindent (i) there exists a common fixed point $\rho\in\fix(\Ti_k)$ for all $k$;\\
\noindent (ii) $\supp(\rho)=\supp(\ker(\Ti)=\supp(\fix(\Ti_k))$ for each $k$. \\
\noindent
Then $\rho'$ is invariant under $\Ti$ only if it is invariant under all $\Ti_k$, that is:
\[ \rho' \in\fix(\Ti) \implies \rho' \in\fix(\Ti_k).\] }

\noindent
{\bf Proof:} 
We show that  $\fix(\Ti)=\bigcap_k\fix(\Ti_k)$. First, linearity of $\Ti$ implies that 
$\fix(\Ti)\geq \bigcap_k\fix(\Ti_k)$. The hypotheses are needed to show $\fix(\lind)\leq\bigcap_k\fix(\Ti_k)$.
By (ii), $\rho$ is a maximal-rank state in $\fix(\Ti)$ and $\rho\in\fix(\Ti_k)$ for all $k$. Thus,  its 
support $\tilde\Hi$ is invariant for all maps and we can construct reduced maps $\tilde\Ti_k$ as we did in 
Theorem \ref{thm:dualkernel-general} for reduced generators $\tilde\Li$. The reduction of $\rho$ to this support, 
call it $\tilde\rho,$ is full-rank. In view of (ii), Theorem \ref{thm:dualkernelkraus} implies that
\beqan
 \fix(\Ti)= \rho^\frac{1}{2} ( \fix(\tilde\Ti^\dag)\oplus {\mathbb O} ) \rho^\frac{1}{2} \quad 
\textup{and} 
\quad  \fix(\Ti_k)= \rho^\frac{1}{2} (\fix(\tilde\Ti_k^\dag)\oplus {\mathbb O} )\rho^\frac{1}{2},
\eeqan
for all $k$. Then, by Theorem \ref{thm:commutantkernelkraus}, we also have that
\beqan
\rho^\frac{1}{2} \fix(\tilde\Ti^\dag)\rho^\frac{1}{2} = \rho^\frac{1}{2} \alg\{\tilde\Ti\}' \rho^\frac{1}{2}\quad 
\textup{and} 
\quad  \rho^\frac{1}{2} \fix(\tilde\Ti_k^\dag) \rho^\frac{1}{2} = \rho^\frac{1}{2} \alg\{\tilde\Ti_k\}' \rho^\frac{1}{2}.
\eeqan
\noindent 
Since $\alg\{\tilde\Ti\}\geq \alg\{\tilde\Ti_k\}$ for all $k$, the relevant commutants satisfy 
$\alg\{\tilde\Ti\}'\leq \alg\{\tilde\Ti_k\}',$ for all  $k.$  This inequality may be used to bridge the previous equalities, 
yielding
\begin{center}
\begin{tabular}{ccccc}
$\fix(\Ti)$ & $=$ & $\rho^\frac{1}{2} (\fix(\tilde\Ti^\dag)\oplus {\mathbb O})  \rho^\frac{1}{2}$ & $=$ & 
$ \rho^\frac{1}{2} \alg\{\tilde\Ti\}' \rho^\frac{1}{2} \oplus {\mathbb O}$ \\
&  &  &  &\vspace*{-2mm} \\
&  &  &  & $\text{\rotatebox[origin=c]{-90}{$\leq$}}$ \vspace*{-2mm} \\
&  &  &  &  \\
$\fix(\Ti_k)$ & $=$ &  $\rho^\frac{1}{2}( \fix(\tilde\Ti_k^\dag)\oplus {\mathbb O}) \rho^\frac{1}{2}$  & $=$ & 
$\rho^\frac{1}{2} \alg\{\tilde\Ti_k\}' \rho^\frac{1}{2} \oplus {\mathbb O}$,
\end{tabular} \vspace*{1mm}
\end{center}
for all $k$. From this we obtain $\fix(\Ti)\leq \bigcap_k\fix(\Ti_k)$, which completes the proof.
\qed

\section*{Appendix C: Results and proofs for pure Dicke states}

\noindent 
{\bf Proposition C.1.}  
{\bf (Pseudo-pure Dicke states)}
{\em Pseudo-pure versions of Dicke states are not FFQLS in general}. 

\vspace*{10pt}
\noindent
{\bf Proof:} 
Consider, in particular, the following state of the general form given in Eq. (\ref{rhopp}):
$$ \rho_\text{pp} = (1-\epsilon) \ket{(0011)}\bra{(0011)} + \epsilon \, {\mathbb I}/2^4. $$
To calculate the operator Schmidt span of this state, we first calculate that of the pure state.
The $1|234$-Schmidt decomposition of $\ket{(0011)}$ is
\begin{eqnarray*}
\ket{(0011)}= 
\frac{1}{\sqrt{2}}\Big( \ket{0}\ket{(011)}+\ket{1}\ket{(001)} \Big),
\end{eqnarray*}
with the analogous $123|4$-Schmidt decomposition being obtained by symmetry. 
With the above pure-state Schmidt decomposition, we easily calculate that of the pseudo-pure state:
\begin{eqnarray*}
 \rho_\text{pp}&=& \identity/\sqrt{2}\otimes [\epsilon \ketbra{(001)}+\epsilon\ketbra{(011)}+(1-\epsilon)\identity/8] \\
&&+\epsilon Z/\sqrt{2}\otimes(\ketbra{(011)}-\ketbra{(001)})/\sqrt{2}\\
&&+\epsilon Y/\sqrt{2}\otimes(i\ket{(001)}\bra{(011)}-i\ket{(011)}\bra{(001)})/\sqrt{2}\\
&&+\epsilon X/\sqrt{2}\otimes(\ket{(001)}\bra{(011)}+\ket{(011)}\bra{(001)})/\sqrt{2}
\end{eqnarray*}
Our prescription is to distort the $234$-Schmidt span by $\rho^{-\frac{1}{2}}_{234}$ 
and calculate the generated *-algebra. 
The distortion takes the four Schmidt basis operators above to (after rescaling)
\begin{eqnarray*}
 \identity_{234},\,\,\,\,\,\,\,\,\,\,&\,\,&\ketbra{(011)}-\ketbra{(001)},\\
i\ket{(001)}\bra{(011)}-i\ket{(011)}\bra{(001)},&\,\,&\ket{(001)}\bra{(011)}+\ket{(011)}\bra{(001)}.
\end{eqnarray*}
The *-algebra generated by the span of these operators is $\mathcal{B}(\hilbert_W)\oplus \complex \identity_{\hilbert^{\perp}_W}$, where $\hilbert_W\equiv\textup{span}\{\ket{(001)},\ket{(011)}\}$ and 
$\hilbert^{\perp}_W$ is its orthogonal complement. 
Re-distorting this *-algebra with $\rho^{\frac{1}{2}}_{234}$, we find that it is unchanged by such an action. 
{Further, it is invariant under ${\cal M}_\frac{1}{2}$. 
Appending $\mathcal{B}(\hilbert_1)$, the minimal fixed-point set 
corresponding to the $1|234$ bipartition is
\begin{equation*}
\mathcal{B}(\hilbert_1)\otimes {\cal F}_{\rho_{234}}(\Sigma_{234}(\rho_\text{pp}))=\mathcal{B}(\hilbert_1\otimes\hilbert_W)
\oplus(\mathcal{B}(\hilbert_1)\otimes\identity_{\hilbert_W^{\perp}}).
\end{equation*}
By symmetry, the minimal fixed-point set corresponding to the $123|4$ bipartition is
\begin{equation*}
{\cal F}_{\rho_{123}}(\Sigma_{123}(\rho_\text{pp}))\otimes\mathcal{B}(\hilbert_4)=\mathcal{B}(\hilbert_W\otimes\hilbert_4)\oplus(\identity_{\hilbert_W^{\perp}}\otimes\mathcal{B}(\hilbert_4)).
\end{equation*}
While the two minimal fixed-point sets both contain the target state, they also both contain the identity operator. 
Their intersection is thus the two-dimensional space
\begin{eqnarray*}
 \mathcal{B}(\hilbert_1)\otimes {\cal F}_{\rho_{234}}(\Sigma_{234}(\rho_\text{pp}))\bigcap 
 {\cal F}_{\rho_{123}}(\Sigma_{123}(\rho_\text{pp}))\otimes\mathcal{B}(\hilbert_4)
=\textup{span}\{\identity,\rho_\text{pp}\}, 
\end{eqnarray*}
violating the necessary condition for FFQLS.}
\qed

\vspace*{3mm}

We next present the proof of Proposition \ref{DickeQL}.
The following two lemmas 
are useful to characterize and constrain the intersection of Schmidt spans for generalized Dicke states.

\vspace*{3mm}

\noindent 
{\bf Lemma C.2.} 
{\em Consider a pure state $\ket{\psi}$, neighborhood structure $\{\neigh_j\}$, with $j=1,\ldots,L$,   
and corresponding Schmidt spans $\Sigma_{\neigh_j}(\ket{\Psi})$. For each neighborhood, let $G_j$ be a 
representation of a group which acts trivially on $\Sigma_{\neigh_j}(\ket{\Psi})$. For each such group, we trivially extend its action from 
$\hilbert_{\neigh_j}$ to $\hilbert_{\neigh_j}\otimes\hilbert_{\overline{\neigh}_j}$ by tensor-factoring  
$\identity_{\overline{\neigh}_j}$ to its elements. Then, the group generated by all the neighborhood groups,
\begin{equation*}
G\equiv \langle G_1, G_2, \ldots, G_L\rangle,
\end{equation*}
must act trivially on the intersection $\bigcap_{j}\Sigma_{\neigh_j}(\ket{\psi})\otimes\hilbert_{\overline{\neigh}_j}$.}

\vspace*{10pt}
\noindent
{\bf Proof:} 
Let $v\in\bigcap_{j}\Sigma_{\neigh_j}(\ket{\psi})\otimes\hilbert_{\overline{\neigh}_j}$. Then 
$v\in\Sigma_{\neigh_j}(\ket{\psi})\otimes\hilbert_{\overline{\neigh}_j}$ for all $j=1,\ldots,L$. 
It then follows, by assumption, that the elements of each $G_j$ act trivially on $v$.
\qed

\vspace*{2mm}

\noindent 
\noindent 
{\bf Lemma C.3.} 
{\em
For any generalized Dicke state and a connected neighborhood structure, the intersection of all neighborhood 
Schmidt spans is contained in the symmetric subspace.}

\vspace*{10pt}
\noindent
{\bf Proof:} 
By Eq. (\ref{DickeSch}), it follows that any Schmidt span of a generalized Dicke state is spanned by Dicke states, and hence is contained in the symmetric subspace. By definition, the latter is the trivial representation of the symmetric group. Thus, for any Dicke state, the Schmidt span with respect to some neighborhood is acted on trivially by permutations of that neighborhood's subsystems. For a connected neighborhood structure, the set of neighborhood-wise permutations generates the full  ${\mathcal S}_n$. By Lemma C.2, the intersection of Schmidt spans for that neighborhood structure must be acted on trivially by ${\mathcal S}_n$, whereby the claim follows.
\qed

\vspace*{1mm}

\noindent
{\bf Proposition V.8.}
{\bf (FFQLS Dicke states})
{\em Given $n$ qudits and a connected neighborhood structure $\neigh$, there exists a (non-factorized) FFQLS
generalized Dicke state relative to $\neigh$ if $d(m -1) \geq n$, where $m$ is the size of the 
{\em largest} neighborhood in $\neigh$.}

\vspace*{10pt}
\noindent
{\bf Proof:} 
We show that the entangled $n$-qudit Dicke state $D_{n,m}$ defined in Eq. (\ref{eq:bigdicke}) in 
the main text is FFQLS relative to any connected neighborhood structure containing a 
neighborhood of size (at least) $m$.
As established in Lemma C.3, the assumed connectivity of $\neigh$ implies that the intersection of the 
extended Schmidt spans $\Sigma_{\neigh}( D_{n,m})\otimes\hilbert_{\overline{\neigh}}$ lies in the symmetric subspace.
Thus, it remains to show that any symmetric state not in $\textup{span}( D_{n,m})$ will fail to be included in some extended Schmidt span. Specifically, we show that any such state fails to lie in the Schmidt span corresponding to
a size-$m$ neighborhood, which we label $\neigh_0$.

The candidate FFQLS Dicke state $D_{n,m}$ has the special property that it contains no more than $m-1$ of symbols $0,\ldots,d-1$, and no more than $r$ of the symbol $d-1$. Of all states in the symmetric subspace of $n$ qudits, $D_{n,m}$ is the only one satisfying this property; viewing each label $0,\ldots,d-1$ as a bosonic mode, $D_{n,m}$ constitutes the only possible $n$-particle filling of these modes without exceeding the imposed occupancy limits of $m-1$ (and $r$ for mode $d-1$, respectively).
It remains to show that any symmetric state vector $\ket{\phi}$ not satisfying this mode-occupancy-limit property, and hence not in 
$\textup{span}(D_{n,m})$, fails to be in at least one extended Schmidt span of the state 
$D_{n,m}$. A symmetric state which fails to satisfy the above mode occupancy limits will admit an orthogonal decomposition of the form 
\begin{equation*}
 \ket{\phi}=\alpha\ket{\underbrace{jj\ldots j}_{m}\,}_{\neigh_0}\otimes\ket{q}_{\overline{\neigh}_0}+\beta\ket{r},
\end{equation*}
if a mode $j=0,\ldots,d-1$ is over-occupied, or of the form 
\begin{equation*}
 \ket{\tau}=\gamma\ket{\underbrace{d-1,d-1,\ldots d-1}_{r+1}\ldots}_{\neigh_0}\otimes\ket{s}_{\overline{\neigh}_0}+\delta\ket{t},
\end{equation*}      
if mode $d-1$ is over-occupied. 
The extended Schmidt span $\Sigma_{\neigh_0}( D_{n,m})\otimes\hilbert_{\overline{\neigh}_0}$ is spanned by vectors of the form 
$\ket{\Lambda}_{\neigh_0}\otimes\ket{i}_{\overline{\neigh}_j}$, where $\ket{\Lambda}_{\neigh_0}$ are Dicke states with $m$ symbols drawn (without replacement) from the symbols of $D_{n,m}$, and the $\ket{i}$ form a basis for $\hilbert_{\overline{\neigh}_j}$. Crucially, $D_{n,m}$ is such that 
no $\ket{\Lambda}$ will have more than $m-1$ occurrences of $j$, implying that 
$$\inprod{\Lambda}{jj\ldots j}_{\neigh_0}=0, \quad \quad \inprod{\Lambda}{\underbrace{d-1,d-1,\ldots d-1}_{r+1}\ldots}_{\neigh_0}=0.$$
\noindent 
From this fact, the length of both $\ket{\phi}$ and $\ket{\tau}$ will be decreased when orthogonally projected onto the extended Schmidt span (more precisely, they cannot exceed the lengths $||\beta\ket{r}||<1$ and $||\delta\ket{t}||<1$, respectively). Hence, the states $\ket{\phi}$ and $\ket{\tau}$ are not in the extended Schmidt span $\neigh_0$. This shows that $D_{n,m}$ is the only state vector in the intersection of its extended Schmidt spans, and hence it is FFQLS, as claimed.
\qed

\end{document}